\documentclass[useAMS,usenatbib]{mn2e} 
\usepackage{graphicx}
\usepackage{amssymb,amsmath}
\title[ML for transient discovery in PS1 difference imaging]
{Machine learning for transient discovery in Pan-STARRS1 difference imaging.} 
\author[D.E. Wright \textit{et al.}]{D.E. Wright$^{1}$\thanks{E-mail: dwright04@qub.ac.uk}, 
S.J. Smartt$^{1}$, 
K.W. Smith$^{1}$, 
P. Miller$^{2}$, 
R. Kotak$^{1}$, 
A. Rest$^{3}$,    \newauthor
W.S. Burgett$^{4}$, 
K.C. Chambers$^{5}$,
H. Flewelling$^{5}$,  
K.W. Hodapp$^{5}$,  
M. Huber$^{5}$, \newauthor
R. Jedicke$^{5}$,
N. Kaiser$^{5}$,
N. Metcalfe$^{6}$,
P.A. Price$^{7}$,
J.L. Tonry$^{5}$, 
R.J. Wainscoat$^{5}$ \newauthor 
and C. Waters$^{5}$\\
$^{1}$Astrophysics Research Centre, School of Mathematics and Physics, Queen's University Belfast, Belfast BT7 1NN, UK\\
$^{2}$The Institute of Electronics, Communications and Information Technology, Queen's University Belfast, Belfast, BT3 9DT, UK\\
$^{3}$Space Telescope Science Institute, 3700 San Martin Drive, Baltimore, MD 21218, USA\\
$^{4}$GMTO Corporation, 251 S. Lake Ave., Suite 300, Pasadena, CA 91101, USA\\
$^{5}$ Institute for Astronomy, University of Hawaii, 2680 Woodlawn Drive, Honolulu, HI 96822, USA\\
$^{6}$Department of Physics, Durham University, South Road, Durham DH1
3LE, UK\\
$^{7}$Department of Astrophysical Sciences, Princeton University, Princeton, NJ 08544, USA}

\begin{document}

\date{Accepted ???. Received ???; in original form ???}
\maketitle \label{firstpage}

\begin{abstract} 
Efficient identification and follow-up of astronomical transients is hindered by the need for humans to manually select promising candidates from data streams that contain many false positives. These artefacts arise in the difference images that are produced by most major ground-based time domain surveys with large format CCD cameras. This dependence on humans to reject ‘bogus’ detections is unsustainable for next generation all-sky surveys and significant effort is now being invested to solve the problem computationally.
In this paper we explore a simple machine learning approach to
real-bogus classification by constructing a training set from the
image data of $\sim$32000 real astrophysical transients and bogus detections from the Pan-STARRS1 Medium Deep Survey.  We derive our feature representation from the pixel intensity values of a 20$\times$20 pixel stamp around the centre of the candidates. This differs from previous work in that it works directly on the pixels rather than catalogued domain knowledge for feature design or selection.  Three machine learning algorithms are trained  (artificial neural networks, support vector machines and random forests) and their performances are tested on a held-out subset of 25\% of the training data.
We find the best results from the random forest classifier and
demonstrate that by accepting a false positive rate of 1\%, the
classifier initially suggests a missed detection rate of around 10\%.
However we also find that a combination of bright star variability,
nuclear transients and uncertainty in human labelling means that our
best estimate of the missed detection rate is approximately 6\%.  
\end{abstract}


\section{Introduction} 
\label{sec:intro} 
Current transient surveys such
as Pan-STARRS1 (PS1) \citep{Kaiser10}, PTF \citep{Rau09}, LSQ \citep{Baltay13},
SkyMapper \citep{Keller07} and CRTS \citep{Drake09} are efficient
discoverers of astrophysical transients.  To make these surveys possible
it has become necessary to automate every step in the data pipeline
including data collection, archiving and reduction.  A major goal for
time-domain astrophysics is early detection and rapid follow-up to
enable complete data sets for transients. Artefact rejection has become
the bottle-neck between fast transient detection and our ability to feed
these targets to follow-up surveys such as PESSTO \citep{Smartt13} and 
PTF for early classification.  Current artefact rejection
typically involves deriving some set of parameters from the image data of
individual detections and thresholding each parameter, only promoting
those detections that pass the thresholds to humans for
verification.\\
\indent The numbers of detections that must be scanned by humans is still on
the order of hundreds of objects each night with a high false positive
rate.  The processing artefacts produced are a
result of many factors such as saturated sources, convolution issues and
detector defects amongst others, and to a large extent are common across
all surveys.  For the next generation of survey we cannot expect humans
to remain involved in this process of artefact rejection to the same
extent, where for example we expect on the order of $10^6$ transient
detections per night from LSST\footnote{http://www.lsst.org/lsst/}.\\
\indent Significant effort has been devoted to this problem in
anticipation of these next generation surveys, and to enable rapid turn
around from detection to classification for current surveys. Machine
learning techniques have been used to take advantage of the large
amounts of data gathered by these surveys to train a classifier that can
distinguish real astrophysical transients from artefacts or `bogus'
detections.  Examples include \citet{Donalek08} for the Palomar-Quest
survey, \citet*{Romano06} for SNFactory, and
\citet{Bailey07} and \citet{duBuisson14} for SDSS.  PTF have
demonstrated the ability to efficiently characterise detections and
initiate rapid follow-up, see \citet{Gal-Yam14} for example, where the
problem of real-bogus classification has been addressed by the work of
\citet{Bloom12} and \citet{Brink13}. While these studies do achieve high
levels of performance, the parameters chosen to represent the images are
often dependent on the specific implementation and strategy of the
individual surveys.\\
\indent In this paper we investigate a simple
representation of the images by using the pixel intensities in a region
around a detection in a single difference image.  This choice of
parameterisation is independent of other aspects of the survey, and is
therefore applicable to any survey performing difference imaging while
also lending itself to implementation much earlier in the data
processing pipeline (potentially at the source extraction stage).  We
begin by outlining the real-bogus problem in the context of PS1 in
Section~\ref{sec:problem}, followed by a description of our training set
and image parameterisation in Section~\ref{sec:data}.  In
Section~\ref{sec:optimization} we discuss the various machine learning
algorithms we investigate, outline how we select the optimum classifier,
and report its performance compared with previous work.  We continue in
Section~\ref{sec:furtherAnalysis} with some further analysis to help
understand how we expect the classifier to perform on a live data stream.
 Finally we summarise our results and conclude in
Section~\ref{sec:summary}.

\section{PS1 and the Problem of Real-Bogus Classification}
\label{sec:problem} 
The Pan-STARRS1 system comprises a 1.8\ m primary
mirror  \citep{Hodapp04} and a field-of-view of 3.3\ deg imaged by 60
4800$\times$4800 pixel detectors, constructed from 10\ $\mu$m pixels
subtending 0.258\ arcsec (for more details, see \citet{Magnier13}). The
PS1 filter system consists of 5 filters, g$_{P1}$, r$_{P1}$, i$_{P1}$,
z$_{P1}$ similar to  SDSS griz \citep{York00} with the addition of
y$_{P1}$, which extends redward of z$_{P1}$.  The system is described in
detail by \citet{Tonry12b}. The PS1 Science Consortium (PS1SC) operates
the PS1 telescope performing 2 major surveys. The Medium Deep Survey
(MDS) \citep{Tonry12a} is allocated 25\% of observing time for high
cadence observations of 10 fields, each the size of the PS1
field-of-view. The wide-field 3$\pi$ survey with 56\% observing time
aims to observe the entire sky north of $-$30\ deg. declination with a
total of 20 exposures per year in all five filters for each
pointing.\\
\indent In this paper we use images from the MDS.  Each night 3-5 of the
MDS fields are observed. Each epoch is composed of eight dithered
exposures of 8 $\times$ 113 s in g$_{P1}$ and r$_{P1}$, or 8 $\times$ 240
s in i$_{P1}$, z$_{P1}$ and y$_{P1}$, producing nightly stacked images of
904 and 1632\ s duration \citep{Tonry12a}.  Each stack achieves
5$\sigma$ depths of around 23.3 mag in g$_{P1}$, r$_{P1}$, i$_{P1}$,
z$_{P1}$ and 21.7\ mag in y$_{P1}$.  Images from the PS1 system are processed by the Image Processing
Pipeline (IPP; \citet{Magnier06}), on a computer cluster at the Maui
High Performance Computer Center (MHPCC). The images are passed through
a series of processing stages including device ‘detrending’,  masking
and artefact location. Detrending includes bias correction and
flat-fielding using white light flat-field images from a dome screen, in
combination with an illumination correction obtained by rastering
sources across the field-of-view. After deriving an initial astrometric
solution, the flat-fielded images are then warped onto the tangent
plane of the sky using a flux-conserving algorithm. The plate scale for
the warped images was originally set at 0.200 arcsec pixel$^{-1}$, but
has since been changed to 0.25 arcsec pixel$^{-1}$ in what is known
internally as the V3 tessellation for the MDS fields. Bad pixels are
masked on the individual images and carried through the stacking stage
to give the ‘nightly stacks’.\\
\indent Difference imaging is performed on a daily basis by two
independent pipelines.  IPP takes the nightly stacks and creates
difference images by subtracting a high-quality reference image from the
new data.  Point spread function (PSF) photometry is then performed on
the difference images to produce catalogues of variables and transient
candidates \citep{Gezari12, McCrum14}.  The Transient Science
Server (TSS) developed by the PS1SC ingests catalogues of detections of
residual flux in the difference images and presents potential transients
for human eyeballing.\\ 
\indent In parallel, an independent set of
difference images are produced at the Centre for Astrophysics at Harvard
from the nightly stack images using the PHOTPIPE \citep{Rest14, Rest05}
software. A custom-built reference stack is produced and subtracted from
the IPP nightly stack to produce an independent difference image. This
process is described in \citet{Gezari10}, \citet{Gezari12},
\citet{Chomiuk11}, \citet{Berger12}, \citet{Chornock13} and
\citet{Lunnan13}, and potential transients are visually inspected for
promotion to the status of transient alert. A cross-match between the
TSS and the PHOTPIPE transient streams is performed and agreement
between the detection and photometry is now excellent, particularly
after the application of uniform photometric calibration based on the
`ubercal’ process \citep{Schlafly12, Magnier13}. \\

\subsection{Artefacts in Difference Imaging} 
\label{sec:artefacts}
\begin{figure*}
   \includegraphics[width=164mm]{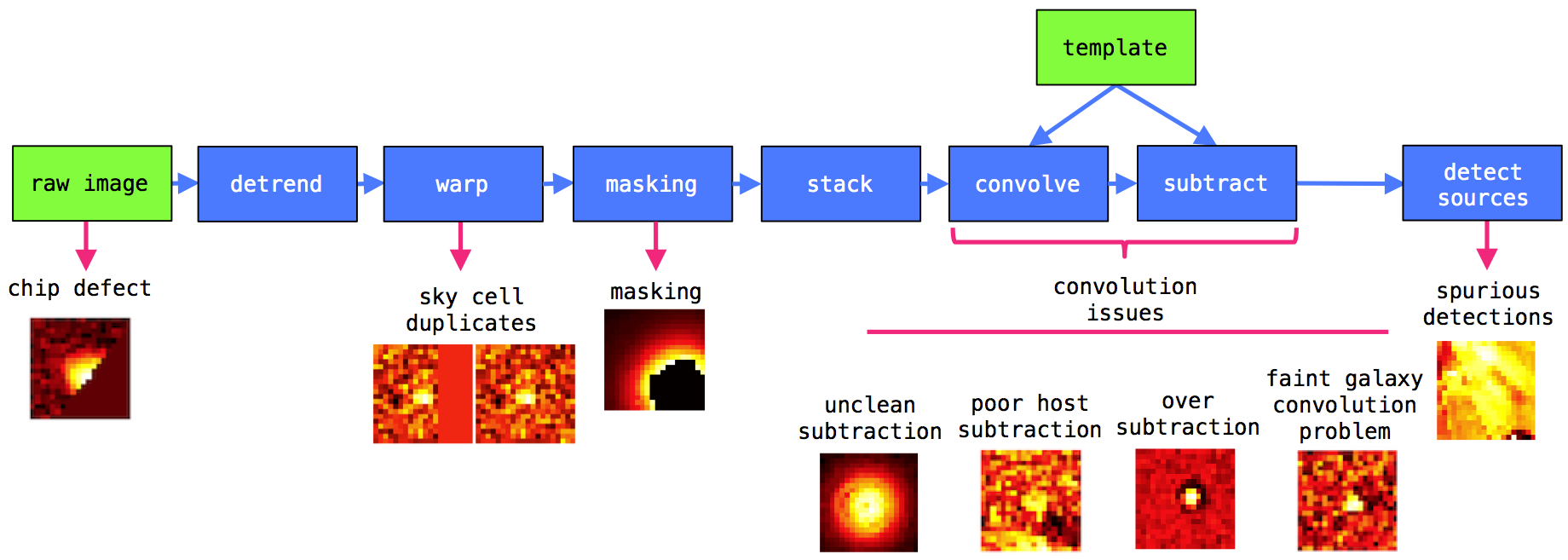}
   \caption{Modular diagram of the IPP difference imaging steps and
     the types of artefacts arising from each stage.} 
   \label{fig:IPP} 
\end{figure*}

\indent  In this work we only use detections from IPP difference imaging and not the
independent PHOTPIPE detections.  In Fig.~\ref{fig:IPP} we show a modular
diagram of IPP difference imaging process and the sources of the main types of
artefact.\\
\indent The first source of bogus detections are chip defects, which
take various forms.  After detrending the chip data are resampled and
geometrically warped to fit a unit area of sky that the data are
projected onto, known as a sky cell.  Occasionally a transient will lie
on a region of the detector that when projected onto the sky falls on
overlapping sky cells.  This results in duplicate warp images of the
same chip data, with the object lying close to one of the skycell edges.  After warping,
sky cell edges, chip defects and saturated sources are masked.  Masked
pixels in individual exposures are propagated through the stacking stage.\\
\indent A kernel is derived to degrade a high quality template image
to match the nightly stack.  The template is convolved with this
kernel and subtracted from the nightly stack.  This series of steps
leads to a class of artefacts which we refer to as convolution
issues.  In general these arise from the derived kernel not being
able to accurately match all sources in the template to those in the
nightly stack.  This causes problems with bright sources where the
kernel is unable to fit the entire PSF of the detection in the nightly
stack image.  These artefacts appear as high signal-to-noise (S/N) PSFs but with darker rings appearing in the
wings, an example is shown in the bottom panel of
Fig.~\ref{fig:bogusVectors}.  We call these unclean subtractions.  The flux in
these detections is probably due to a bright stellar variable. Identifying variable stars (and AGNs) is quite a 
different problem to detecting transients and we have chosen not to 
try to tailor our algorithms to do both. The efforts in this paper are
focused on finding transients, although inevitably stellar variables from
very faint host stars are detected.  Hence we discard these bright stellar
variables that appear in the difference images as they are straightforward
to identify. We find these detections make up $\sim$10\% of the bogus 
detections.\\
\indent The same convolution issues can lead
to poor host galaxy subtraction, where an inadequately convolved host
can be over or under subtracted leaving a pattern of positive
and negative flux.  This makes it difficult to disentangle any potential real
detections.  The third convolution issue we highlight in
Fig.~\ref{fig:IPP} arises when point-like sources in the template image are 
broader than that of the nightly stack resulting in an over subtraction in the wings of the source in
the difference image.  This happens when observing conditions
have been particularly good and the nightly stack is of higher quality than
the template image (this is not a frequent occurrence).  The final artefacts from the convolution and
subtraction stage are convolution
problems in the cores of faint galaxies, manifesting themselves as
faint nuclear transients and appearing as positive flux in the
difference image.  Here the convolution step matches the morphology of
the faint galaxy in the template and nightly stacks well, however the
peak flux of the convolved template is lower than that in the nightly
stack.  This results in the nucleus of the faint galaxy being
under subtracted leaving residual flux in the difference image.  These
artefacts are the most difficult to identify by eye but are
distinguished by a narrower PSF than expected.  It is not always clear if the
flux is due to real variability or an artefact of convolution, in any case these 
targets could not be confidently selected as real transients for
follow-up.  This highlights one of the major uncertainties in training
the algorithms --- secure labelling of real and bogus objects, which we return to in
Sections~\ref{sec:contamination} 
and~\ref{sec:missedDetections}.\\
\indent Another source of artefacts arises during the source
extraction phase.  Flux in the nightly stack
from diffraction spikes for example that
have no equivalent in the template image get flagged as potential
transients.  We refer to these as spurious detections in Fig.~\ref{fig:IPP}.\\
\indent Our approach to date for removing these contaminants has been to
attempt to derive a set of filters based on image statistics derived for
each potential transient detection by IPP.  These filters normally take
the form of threshold values for some parameters (see
Section~\ref{sec:flags}). However, the parameter space is typically
large and the work required to manually develop the optimal set of
filters is impractical.  Despite this our current hand-engineered checks
allow only a small fraction of the bogus images through.  This
still produces on the order of a few hundred bogus objects each night
passing the cuts and being presented to human scanners for verification.
 This is approaching the limit of what can comfortably be processed by
humans on a daily basis and clearly a solution needs to be found for the
next generation of survey.\\ 
\indent Over the course of the last $\sim$3
years of the PS1 survey we have accumulated a large amount of data
associated with a few tens of thousands of astronomical sources that have
either been classified as real objects or artefacts using a combination
of the cuts detailed in Section~\ref{sec:flags} and human scanning. 
This readily available data lends itself to data-mining where we hope to
use the historic data to improve on the current method of
real-bogus classification. In Section~\ref{sec:supervised} we outline how
supervised learning can be applied to this archive of PS1 data in order
to construct a real-bogus classifier that can be applied to the nightly
stream of new data gathered from PS1 and future surveys.  First we
describe the cuts we perform.\\

\subsection{Cuts} 
\label{sec:flags}
Prior to ingesting detections from IPP difference imaging into a MySQL
database at Queen's University Belfast (QUB) we perform pre-ingest cuts based on the detection of saturated, masked or suspected
defective pixels within the PSF area. Taking as a typical night 3rd
September 2013 (56548 MJD (Modified Julian Date)), the 7 nightly stacks
produced 366267 detections ($\sim$52000 detections per stack), the
pre-ingest cuts rejected 94.88\% of these detections.\\
\indent  The $\sim$18750
detections passing the pre-ingest cuts are associated with transient
candidates if there are two or more quality detections within the last
seven observations of the field, including detections in more than one
filter, and an rms scatter in the positions of $\leq$0.5\ arcsec. Each
quality detection must be of more than 3$\sigma$ significance
and have a Gaussian morphology (XYmoments $<$1.2). These post-ingest
cuts also include checks for convolution issues, proximity to bright
objects and `NaN' values close to the centre of bright PSFs.  63\% of the
detections that passed the pre-ingest cuts were rejected during the
post-ingest cuts.  The remaining detections were promoted for human
screening, where 37\% of the detections were deemed to be real.  These
real transient candidates are cross-matched with catalogues of
astronomical sources in the MDS fields.  We use our own MDS catalogue
and also extensive external catalogues (e.g. SDSS, GSC, 2MASS, NED,
Milliquas\footnote{http://quasars.org/milliquas.htm}, Veron AGN, X-ray
catalogues) to make a contextual classification of supernova, variable
star, active galactic nuclei or nuclear transient.  We also cross-match
with the Minor Planet Centre to reject asteroids, though most are
removed during the construction of the nightly stacks.

\subsection{Supervised Learning for Classification}
\label{sec:supervised} 
\begin{figure}
   \includegraphics[width=41mm]{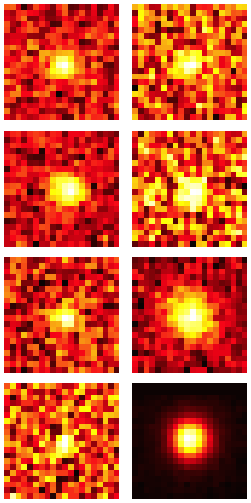}
   \includegraphics[width=41mm]{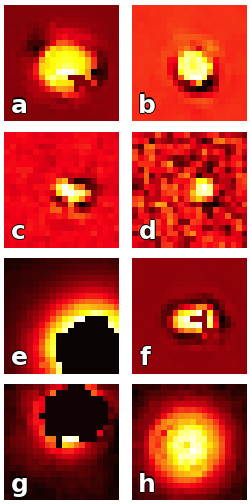} 
   \caption{Example detections randomly
   selected from the training data.  The 2 columns on the left show
   examples labelled as real and the 2 on the right show those labelled
   as bogus.  Bogus detections \textbf{a-c} and \textbf{f} show signs of over subtraction,
   with \textbf{a} and \textbf{f} also showing masking.  \textbf{d} is a faint galaxy
 convolution problem. Detections \textbf{e} and \textbf{g} are
 saturated sources that have been masked.  Finally, \textbf{h} is an example of
an unclean subtraction of a bright star.} 
   \label{fig:trainingExamples} 
\end{figure}

In general supervised learning entails learning a
model from a training set of data for which we provide the desired
output for each training example.  For the purposes of designating a
detection as a real transient or a processing artefact, the desired output
for each image is discrete.  In such cases the problem is a supervised
classification task for which there are a vast array of machine learning
algorithms.  In Section~\ref{sec:optimization} we discuss the algorithms
we try; however all such algorithms are trying to learn a model from the
training data that will allow them to map the input parameterisation of
each training example (see Section~\ref{sec:features}) to the desired
output or \textit{label}, while at the same time ensuring the model
performs well on data not seen during the training phase.  For building
a real-bogus classifier this is an obvious avenue to pursue as we have a
large sample of historical data for which we have labels provided by our
current cuts and also through human eyeballing.\\ 
\indent In Fig.~\ref{fig:trainingExamples} we show a sample of both real and bogus
examples drawn at random from the training data.  Often bogus
detections show a combination of the factors we describe in
Section~\ref{sec:artefacts} and typically this affects the centroiding
during the source detection stage.

\section{Training Set and Feature Representation} 
\label{sec:data} 
As discussed in the previous section we must provide a labelled training
set from which the classifier can learn to recognise the characteristics
that can identify detections as being members of one of the classes: real or
bogus.  In order to learn a model that will generalise well to detections
in new observations, it is important that detections in the training set
are representative of all detections we expect to see.  In practice this is
easiest to achieve by providing the learning algorithm with the largest
possible training set, indeed \citet{Brink13} attribute much of their
improvement in performance over \citet{Bloom12} to using a training set
with two orders of magnitude more training examples.  In the remainder
of this section we describe the compilation of the training set, starting
with a description of our training example selection process and
labelling.

\subsection{Training Set} 
\label{sec:trainingSet} 
Over the past 3 years $\sim$1 million potential transients have been
catalogued in the MDS by the TSS.  Approximately 8000 of these objects have been selected by humans as real transients and
promoted as potential targets for spectroscopic follow-up.  As of the
end of the survey in May 2014, 515 transients had spectroscopic classifications.\\
\indent
The aggregate catalogue information for all objects extracted by IPP and
which pass the pre-ingest cuts described in
Section~\ref{sec:flags} are stored in a database at QUB.  Individual detections are associated with an
object if they are spatially coincident within 0.5 arcsec.  This
information is presented to humans in the form of
webpages\footnote{Similar webpages are made public for the PS1 3Pi
Survey at: http://star.pst.qub.ac.uk/ps1threepi/psdb/public/}.  The
webpages show all the photometric points produced by IPP in a
multi-colour lightcurve. The number of photometric detections
typically ranges from a few to a few dozen depending on the magnitude
and timescale of the transient objects (see \citet{Rest14},
\citet{McCrum14}, \citet{Chomiuk11} for examples of lightcurves). These webpages also present a subset of the image postage-stamps of the detections associated with an object (target image, reference image and difference image). 
This subset contains the first detections of the object of which there
are always at least 2 (see Section~\ref{sec:flags}) and up to 5
subsequent detections.  Each object is then eyeballed by a human, those that appear to be real
transients are promoted as potential targets
for scientific follow-up, while artefacts are discarded.\\
 \indent Our training examples are drawn from the subset
 of detections we choose to present on the human digestible webpages for each object, as detailed above (typically 2-3
 but less than 7).  The majority of real examples were taken from
detections of promoted objects with no spectroscopic classification.  There is no
guarantee that all detections of a promoted object are necessarily
a result of good image subtractions.  This prohibits simply assigning a
label of real to all individual detections associated with a promoted target.  In order to
to ensure that we have a secure, reliable and clean set of real detections for training, we inspected and individually labelled 4352 detections
(from 1919 different transients) as real,
discarding any artefacts from the training set.  We augmented this
sample of real detections with data from 53 spectroscopically
confirmed supernovae (from Dec. 2012 to Jan. 2014) for which we used the complete set of
detections ($\sim$31 detections per object on average).  These were again manually checked to remove bogus
detections.  We held out the first detections of all 53 SNe, which we
use for testing in Section~\ref{sec:earlydetection} and all detections
of PS1-13avb, which we use in Section~\ref{sec:lightcurve}.  This leaves an additional 1603 real training examples
bringing the total to 5955 real detections.\\
\indent Over the course of the survey approximately 800000 objects
have been discarded as artefacts providing on the order of $10^6$
examples of bogus detections.  We randomly sample
from the available bogus examples and aim for 4 times more bogus
examples as real, this is similar to the proportions used by
\citet{Brink13}.  Initial tests with classifiers showed that a
significant proportion of the false positives appeared to be clean
subtractions.  We improved the purity of the bogus sample by examining
the randomly selected bogus detections and added any detections that looked like
real transient subtractions to the list of real examples (the effect of label
contamination is further discussed in Section~\ref{sec:contamination}).
This produced an extra 464 examples for the set of real detections resulting
in a final total of 6419.  We then selected 4 times as many bogus images from
the remainder of the bogus examples we inspected, producing a sample of
25676 bogus detections.\\
\indent The final training set contains 32095
training examples.  We divide the training examples into 2 sets,
distributed as follows; 75\% for training and cross validation, and 25\%
for testing.  The training examples are randomly shuffled prior to
splitting with the caveat that all detections on the same night of
a given object are included in the same set.  This is to avoid detections
with almost identical statistics being in multiple sets and giving a
false impression of a classifiers performance.  The construction of the
data set is summarised in Table~\ref{tab:datasets}.  The label for each
training example is a 1 or 0, with 1 representing a label of real and 0
bogus.\\ 
\indent The training set we have constructed is representative,
containing examples of detections from different chips, seeing conditions,
and filters, with various levels of S/N and examples
of all types of processing artefact.

\begin{table} \centering
    \caption{Composition of data sets.} \begin{tabular}{lccccccccc}
    \hline Set   &&& Real &&& Bogus &&& Total\\ \hline \hline Training
    &&& 4800 &&& 19271 &&& 24071\\ Test &&& 1619 &&& 6405 &&& 8024 \\
    \hline Total &&& 6419 &&& 25676 &&& 32095\\ \hline \end{tabular}
    \label{tab:datasets}
\end{table}

\subsection{Feature Representation} 
\label{sec:features} 
Machine learning algorithms require a 1-dimensional (1-D) vector representation
of each training example, where each element of the vector corresponds
to some numeric data or \textit{feature} that may be
useful to the algorithm for discerning examples belonging to each class.
 Previous work in the area of real-bogus classification, has focused on
using parameters contained in catalogues generated by the processing
pipeline and more complex features derived from that information to
represent the detections, see Table 1 from \citet{Brink13} and Table 1 from
\citet{Romano06}.\\ 
\indent The catalogue features available to
individual surveys depend on the implementation of their image
processing pipeline.  When applying machine learning for real-bogus
classification to a new survey it may not be possible to calculate these
features based on the information available in the catalogues.  There is
also potential to spend a lot of time deriving and testing ways to
combine the catalogue information that is available into features that
we hope capture the differences between real and bogus detections.  Bogus
detections are the result of many factors and establishing a set of features
that can encapsulate them all is difficult.  In contrast simply
representing the detections by their pixel intensity values requires no time
spent developing or tuning feature extractors.  Previous work that
relies solely on the pixel data has proven effective for simple visual
classification tasks, such as hand written digits \citep{LeCun98}.  For
more complex tasks or to boost performance much of this work has been
performed by learning a hierarchy of unsupervised features from the pixel
data (\citet*{Coates11}; \citet{LeCun98}). Establishing a firm benchmark
on the pixel intensity representation allows us to assess the potential
gains from applying these more complex methods and is the main focus of
this paper.  Using this representation we expect the learning algorithm
to identify salient relationships between pixels for the classification
task.  In the next section we discuss our choice of features and
continue in the following section by describing the preprocessing steps
we apply before training.

\subsubsection{Feature Vector Construction} 
\label{sec:vectors} 
\begin{figure}
   \includegraphics[width=84mm]{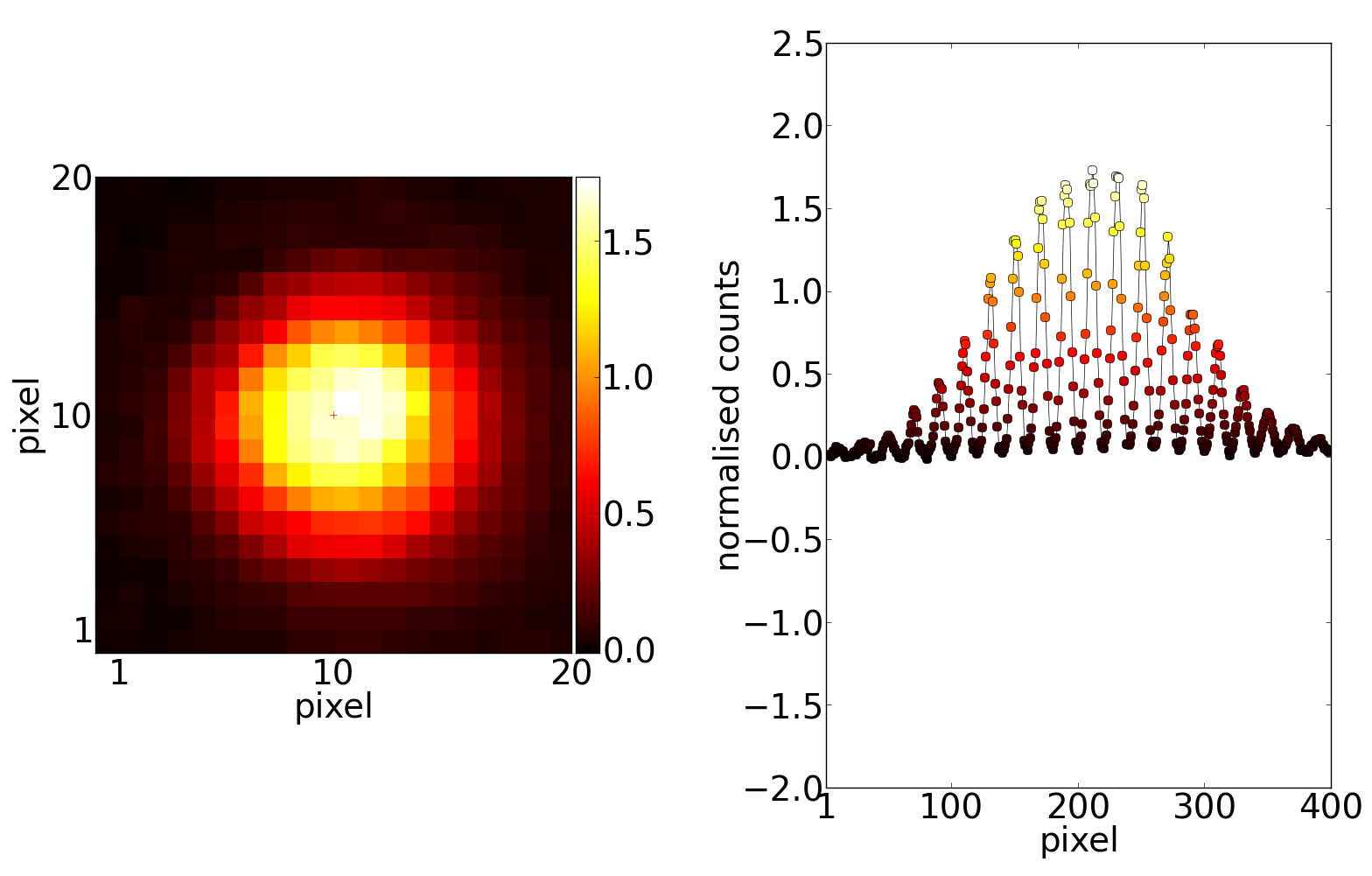}
   \includegraphics[width=84mm]{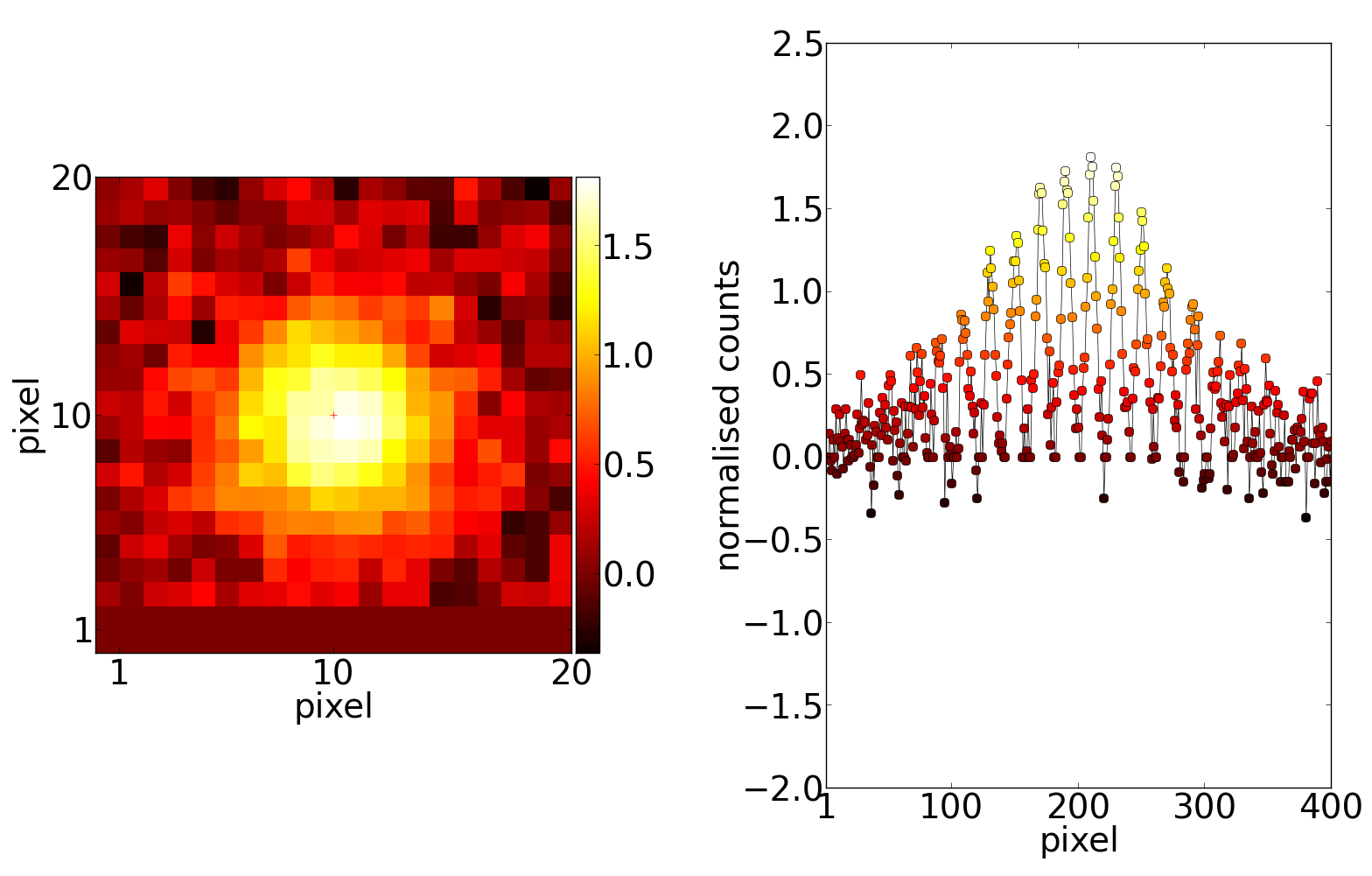}
   \includegraphics[width=84mm]{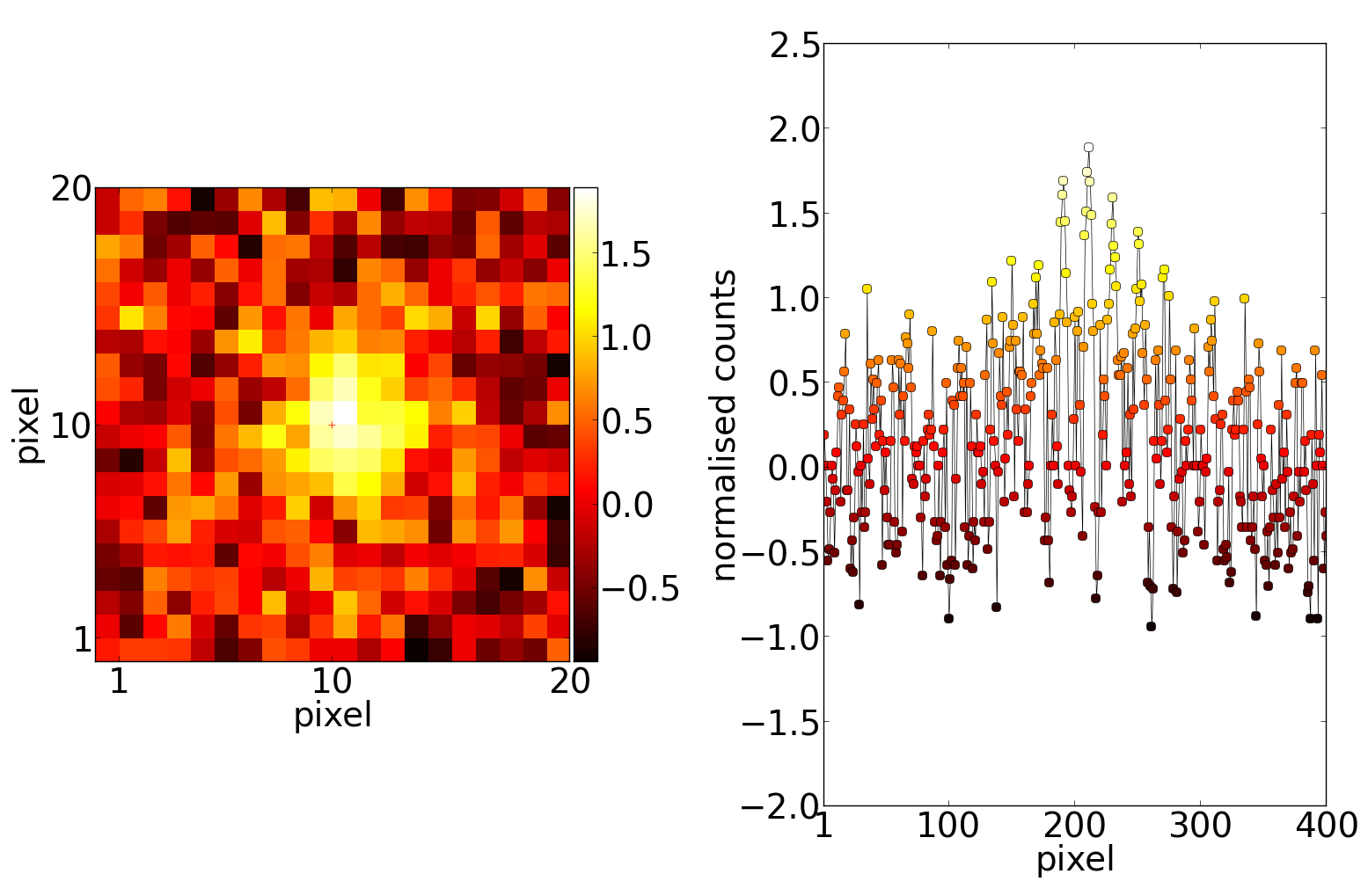}
   \includegraphics[width=84mm]{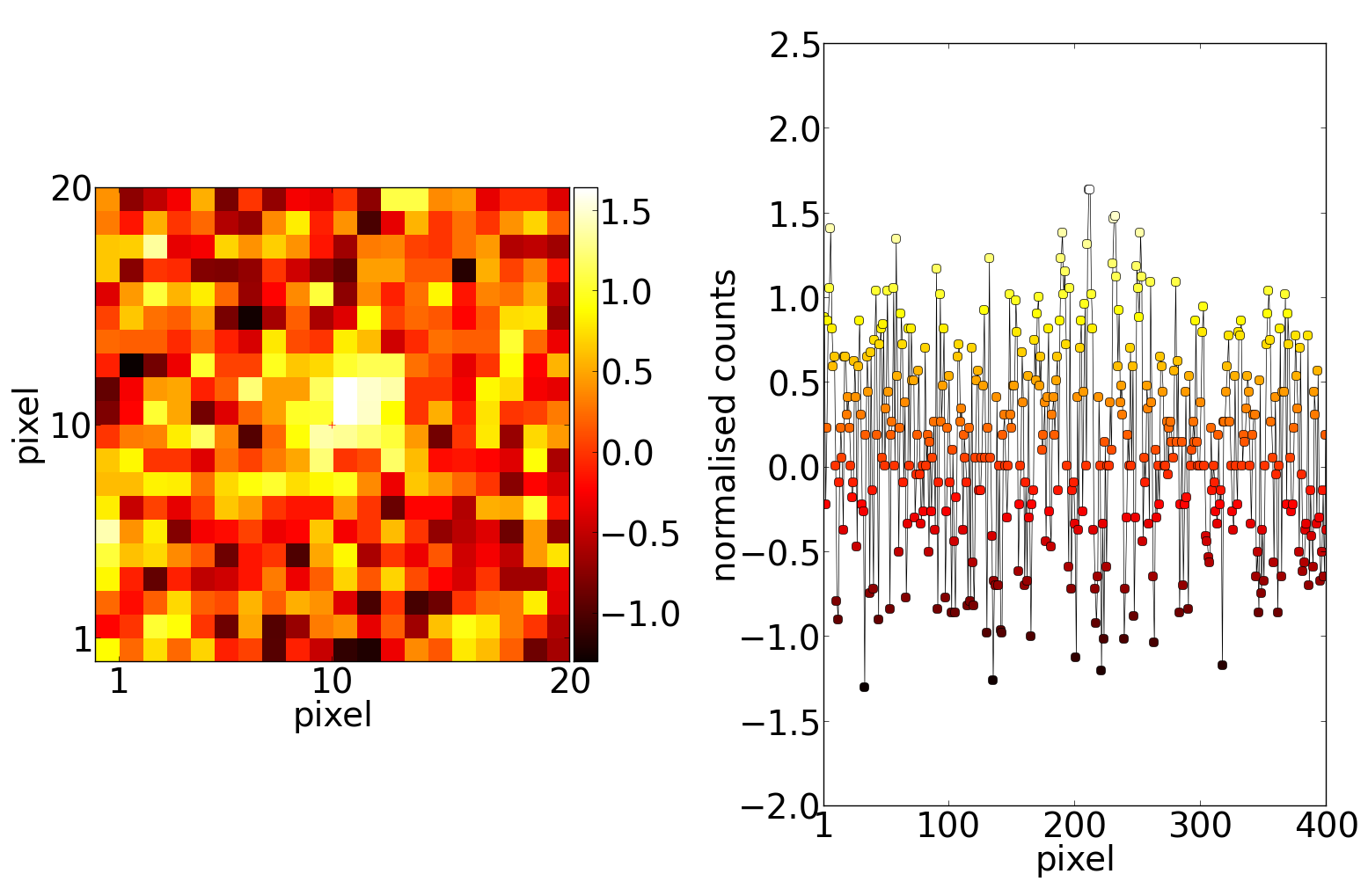} 
   \caption{Visualisation of feature vectors for detections labelled as real.  The feature vectors are
   constructed by shifting off each column of the 20$\times$20 pixel
   substamp on the left and appending them together to produce the 400
   element 1-D feature vector depicted on the right.}
   \label{fig:realVectors} 
\end{figure}

\begin{figure}
   \includegraphics[width=84mm]{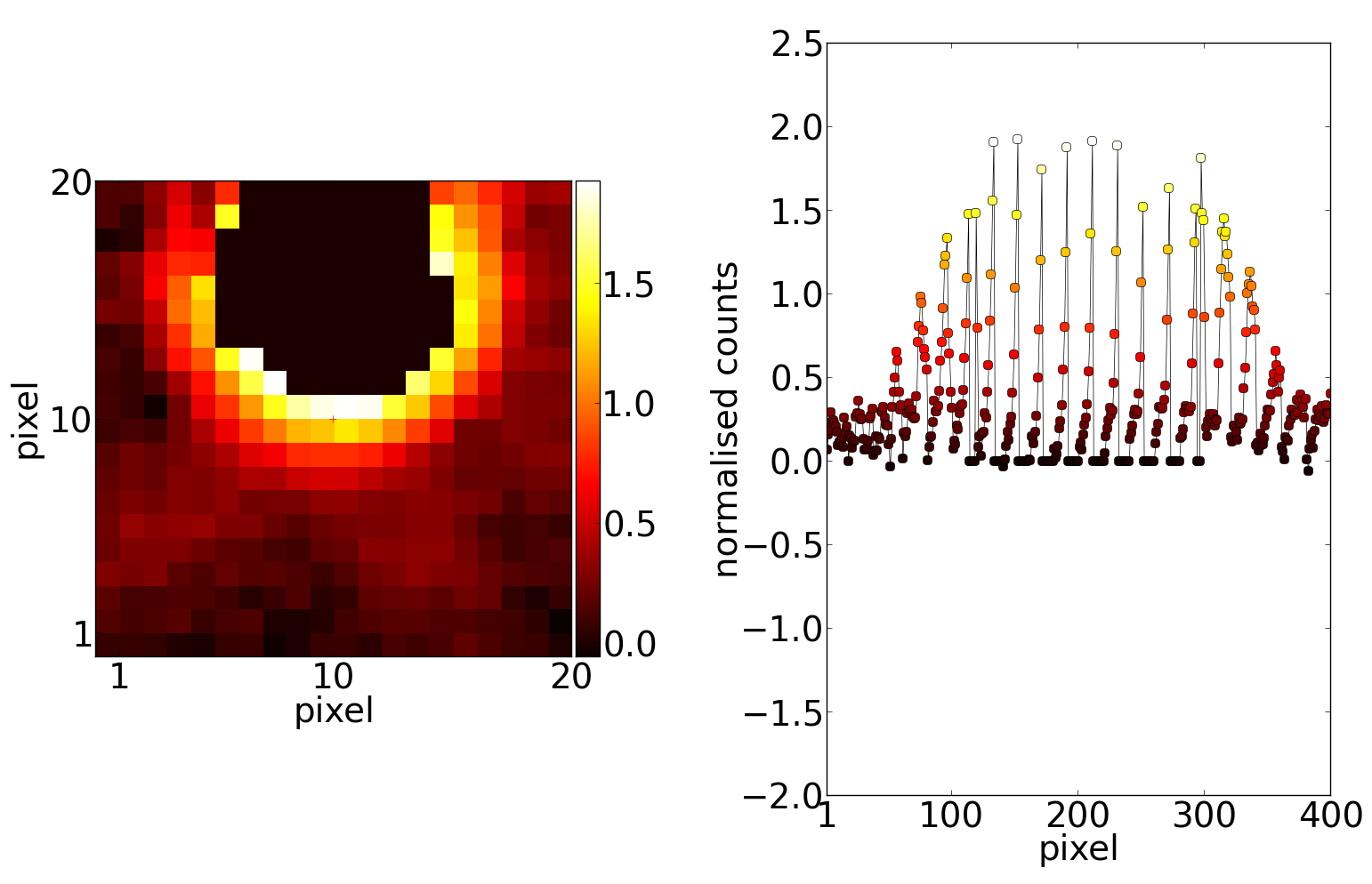}
   \includegraphics[width=84mm]{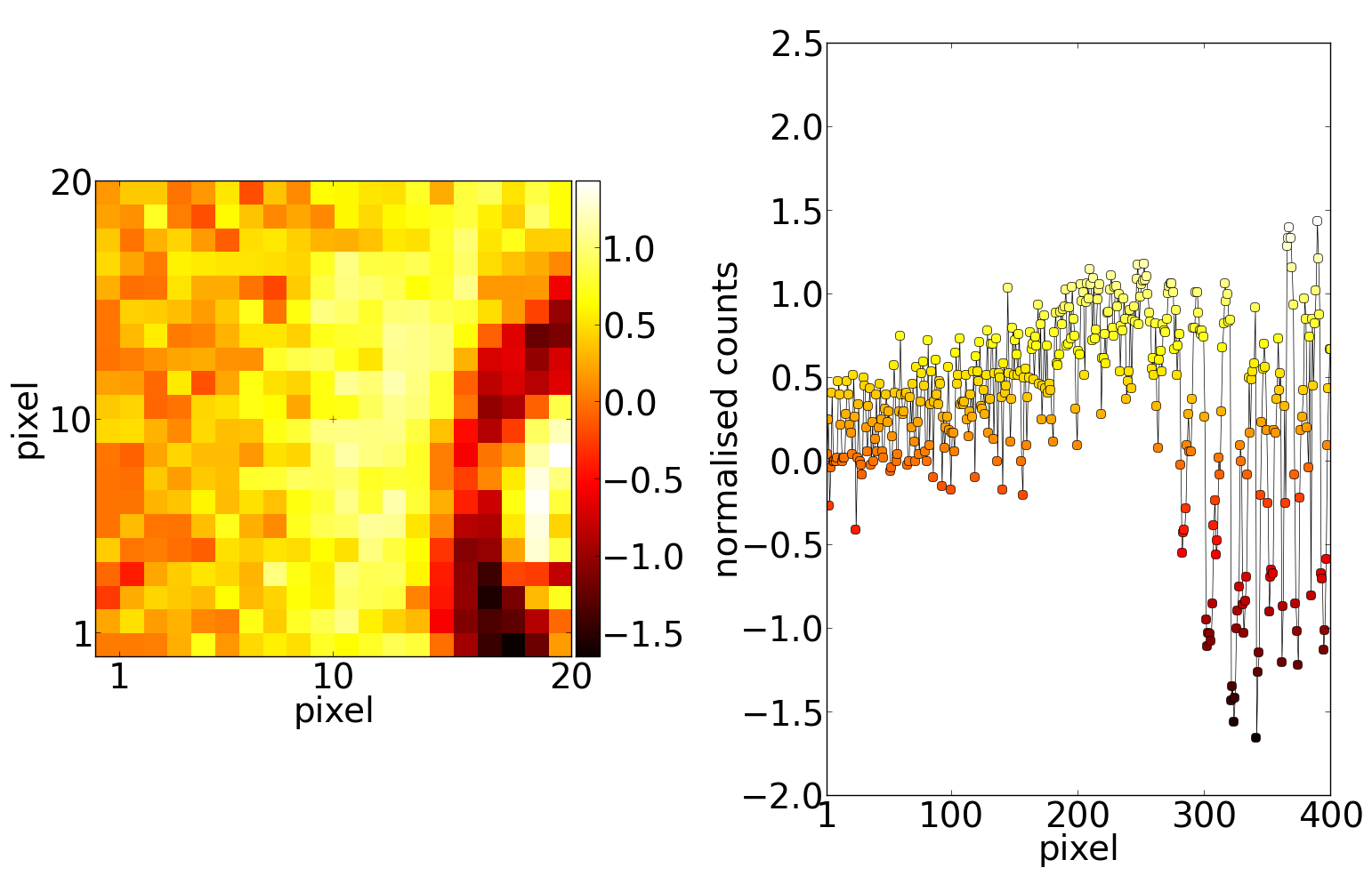}
   \includegraphics[width=84mm]{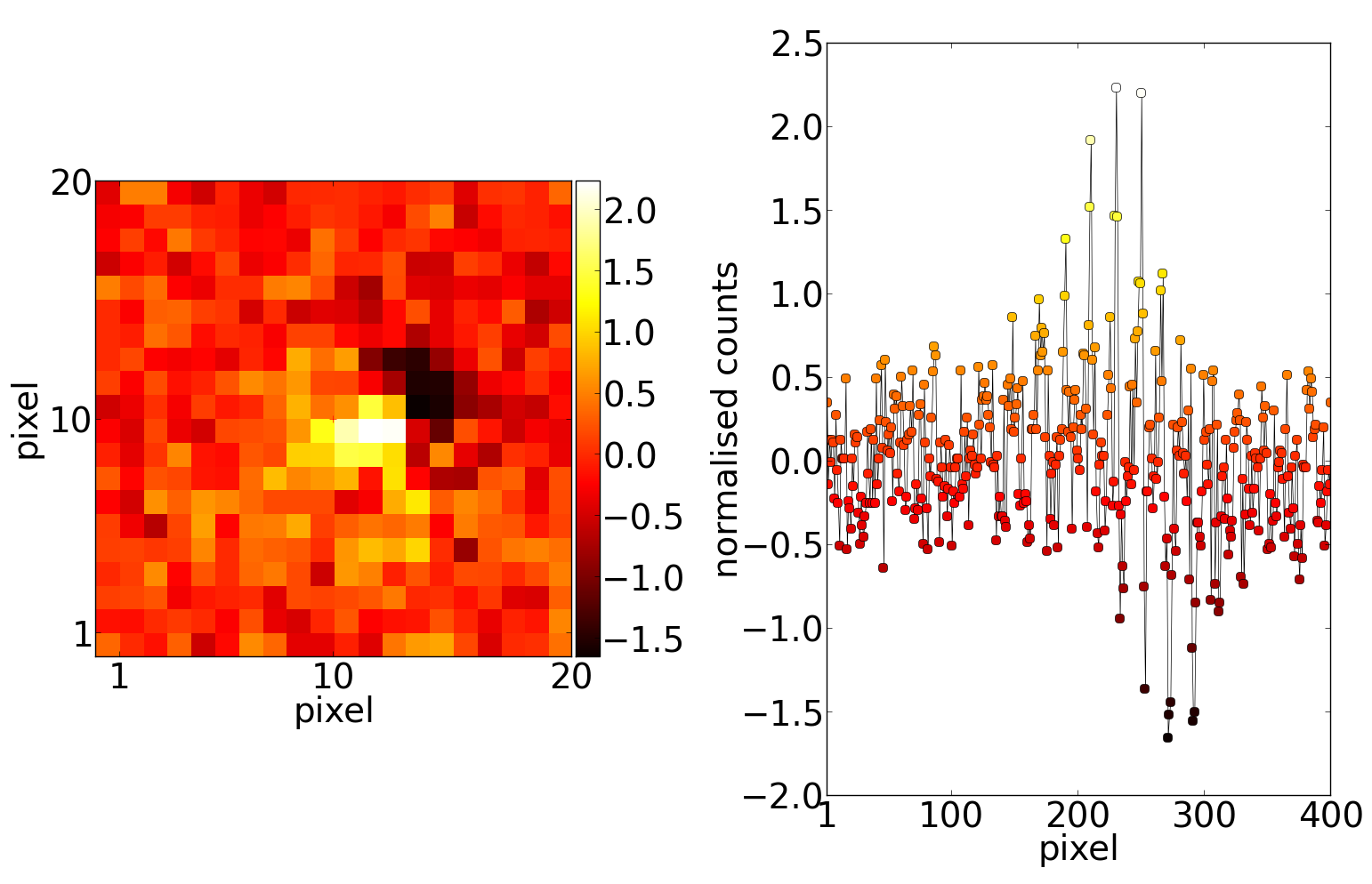}
   \includegraphics[width=84mm]{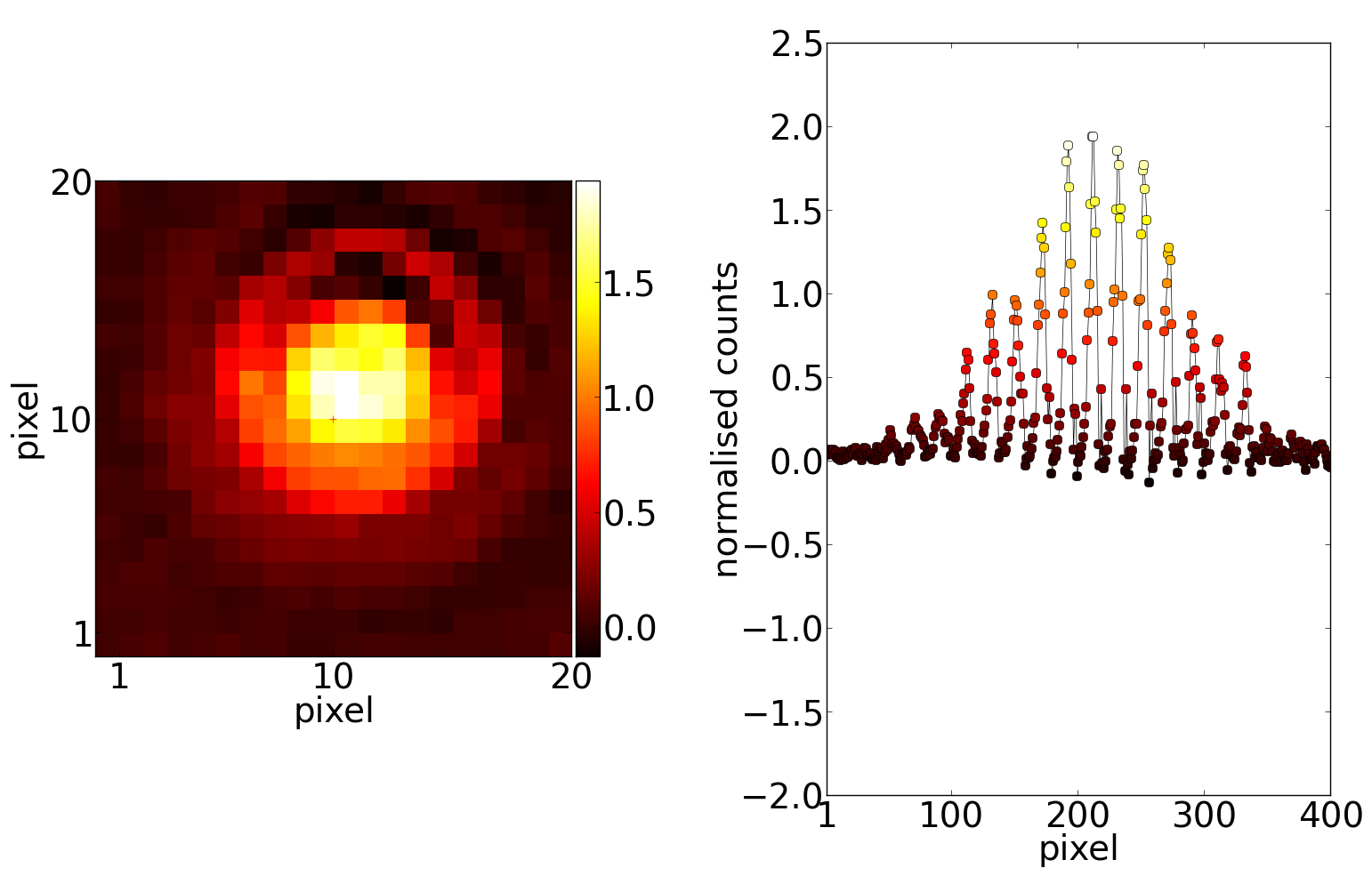} 
   \caption{Similar to Fig.~\ref{fig:realVectors} but for bogus examples.}
   \label{fig:bogusVectors} 
\end{figure}

To
represent our training examples, we use the pixel data itself.  For a
given training example, we construct its feature vector by selecting a
20$\times$20 pixel area (corresponding to $\sim$5 times the average
seeing of PS1) around the centre of what IPP considers a transient,
which we refer to as a substamp.  The
1-D vector is constructed by shifting off each column of the substamp and
concatenating those columns together to produce a 400 element vector of
pixel intensity values.\\ 
\indent In Fig.~\ref{fig:realVectors} we show
visualisations of these feature vectors along with the substamp from which
they were constructed for examples of real detections and for various levels
of S/N.  In Fig.~\ref{fig:bogusVectors} we show detections labelled as bogus
with examples of different types of artefact.  A learning algorithm will
learn to identify patterns in the feature vectors that are characteristic
of examples belonging to the two classes.\\ 
\indent The choice of
feature representation is independent of the implementation of the rest
of the image processing pipeline and survey, with the assumption that
the pixel level data is easily accessible.

\subsubsection{Feature Preprocessing} 
\label{sec:preprocessing} 
Aside
from the image processing steps carried out by the pipeline, we carry
out 2 additional transformations of the data.  We first replace any
`NaN' pixel values with 0s.  `NaN' pixel values typically arise from
masking or floating point overflows during image processing.  We choose
to replace these pixel values with 0 so as not to influence the next
step in the preprocessing phase.  As a second step we apply a feature
normalisation function which allows classifiers to focus on relative
pixel intensities and limits the effect of absolute brightness on the
classifiers.  We apply the following normalisation: 
\begin{equation}
f(x) = \frac{x}{|x|}\ log\left(1 + \frac{|x|}{\sigma}\right)
\end{equation} 
\noindent where $x$ is a feature vector and $\sigma$ is
the standard deviation of the pixel intensity values for that feature
vector.  This is the same normalisation function used by
EyE\footnote{http://www.astromatic.net/software/eye} \citep{Bertin01EyE}
and similar to that of \citet{Romano06}.

\section{Optimisation of the Classification System}
\label{sec:optimization} 
In order to achieve the best performance from
the machine learning algorithms discussed in the following sections it
is necessary to optimise the hyperparameters of each.  This is done by a
process known as cross validation which is a brute force search of the
hyperparameter space, where a model is trained with the hyperparameters
selected at predetermined intervals within the space.  The best
combination is selected by measuring the performance in a held out
sample of the 24071 training examples.\\ 
\indent Below we give a brief
introduction to each of the classifiers.  We also point out the free
parameters that must be selected by cross validation and discuss this
process in depth in Section~\ref{sec:validation}.  To end this section
on optimisation we show the performance of each classifier on the
out of sample data in the test set.

\subsection{Artificial Neural Networks} \label{sec:nn} Artificial Neural
Networks (ANN) comprise a number of interconnected nodes arranged into a
series of layers.  In this study we limit ourselves to a 3-layer ANN
(consisting of an input layer, a hidden layer and an output layer) as
those with more than one hidden layer need more careful training and
require more computational power \citep*{Hinton06}.  For our purposes we
train feed-forward ANNs with back-propagation and randomly initialised
weights, where the activation of each node is calculated with the
logistic (sigmoid) function.\\ 
\indent By limiting many of the choices
for the structure of the ANNs we remove the need to select these
hyperparameters during the cross validation phase in
Section~\ref{sec:validation} which significantly reduces the complexity
of the space we have to search.  This economy of computation comes at
the cost of not testing regions of the parameter space (e.g. other
activation functions) and restricting the representational power of the
ANNs by requiring a single hidden layer.  We are however left with only
2 hyperparamters to choose namely, the number of nodes that make up the
hidden layer, $s_2$ and the regularisation parameter, $\lambda$ through
which we attempt to prevent overfitting.  There is some suggestion
\citep{Murtagh91, Geva92} that the optimal number of nodes in the hidden
layer ($s_2$) is $2n+1$, where n is the number of input features.  In
our case  $n$ is fixed at 400 input features, suggesting we should train
ANNs with $s_2=801$ nodes, however training such large networks is
beyond the scope of this work and we instead choose to test values of
$s_2$ in the range 25-200.\\ 
\indent We use our own vectorised
implementation of ANNs written in
python\footnote{https://www.python.org}.  The code relies on
numpy\footnote{http://www.numpy.org} for efficient array manipulations
and scipy\footnote{http://docs.scipy.org/doc/scipy/reference/index.html}
for optimisation of the objective function.

\subsection{Random Forests} 
\label{sec:rf} 
Random Forests (RFs) aim to
classify examples by building many decision trees from bootstrapped
(sampled with replacement) versions of the training data
\citep{Breiman01}.  Classifications are then assigned based on the
average of the ensemble of decision trees.  Each individual tree is
grown by randomly sampling $k$ features from the $n$ input features and
selecting the feature that best separates real examples from bogus as
informed by the gini function.  We use
scikit-learn's\footnote{http://scikit-learn.org/stable/index.html}
implementation of RFs where we select hyperparameters by assigning
values to variables \texttt{n\_estimators}, \texttt{max\_features} and
\texttt{min\_samples\_leaf}; the total number of trees in the ensemble,
the number of features considered at each split and the minimum number
of examples that define a leaf, below which no further splitting is
allowed.  RFs provide the ability to estimate the importance of each
feature which we use in Section~\ref{sec:importance}.

\subsection{Support Vector Machines} 
\label{sec:svm} 
Support Vector
Machines (SVMs) \citep{Cortes95} aim to find the hyperplane in the input
feature space that optimally classifies training examples for linearly
separable patterns, while simultaneously maximising the margin, the
distance between the training examples which lie closest to the
hyperplane, known as the support vectors.  SVMs can be extended to
non-linear patterns with the inclusion of a kernel, where the kernel
transforms the original input data into a new parameter space.  We again
use scikit-learn's implementation of SVMs where we choose the free
parameters namely, the penalty parameter, \texttt{C} (similar to
$\lambda$ for ANNs) and the kernel parameter \texttt{gamma}, which
controls the local influence that support vectors have on the decision
boundary.  We  only try SVMs with a Radial Basis Function
(RBF) kernel, this being the most common choice and again reduces the
parameter space that must be searched.

\subsection{Model Selection} 
\label{sec:modelSelection} 
For each
algorithm discussed above we need a method to choose the optimal
combination of hyperparameters that will achieve the best performance
for the classification task.  In order to compare the relative
performance of the different models we need some Figure of Merit (FoM). 
We use the FoM of \citet{Brink13} which captures the essence of the
problem we are trying to solve.  The FoM is defined as the minimum
Missed Detection Rate (MDR) (False Negative Rate) that gives a False
Positive Rate (FPR) of 1\%.  That is, assuming we are willing to accept
that 1\% of the images deemed real by the classifier and promoted to
human scanners will turn out to be bogus, what fraction of the real
images would be discarded? With this we can select the model that would
discard the least real images while 1\% of images classified as real can
be expected to be bogus.

\subsubsection{Cross Validation} 
\label{sec:validation} 
\begin{figure}
   \includegraphics[width=84mm]{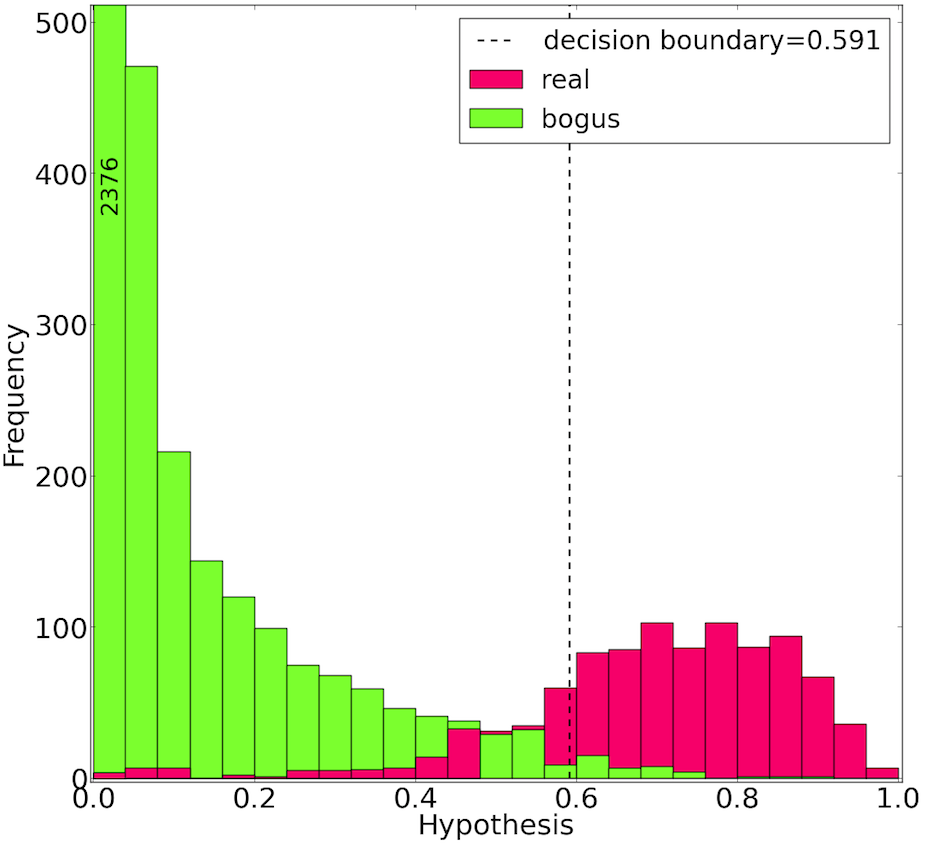} 
   \caption{Hypothesis distribution
   produced during one permutation of 5-fold cross validation.  The
   hypotheses shown are for images in the held-out fold.    Green shows
   the hypotheses for the validation examples labelled as artefacts and
   red those labelled as real.  The decision boundary is selected such
   that the fraction of detections labelled as bogus lying above the
   decision boundary is 0.01.  The False Positive Rate can be visualised
   as the fraction of green bars with a prediction greater than the
   decision boundary, the MDR is the fraction of red bars with
   predictions less than the decision boundary.  The first interval has a frequency of 2376, but the plot is 
truncated for clarity.} 
   \label{fig:nnHypo}
\end{figure}

\begin{figure}
   \includegraphics[width=84mm]{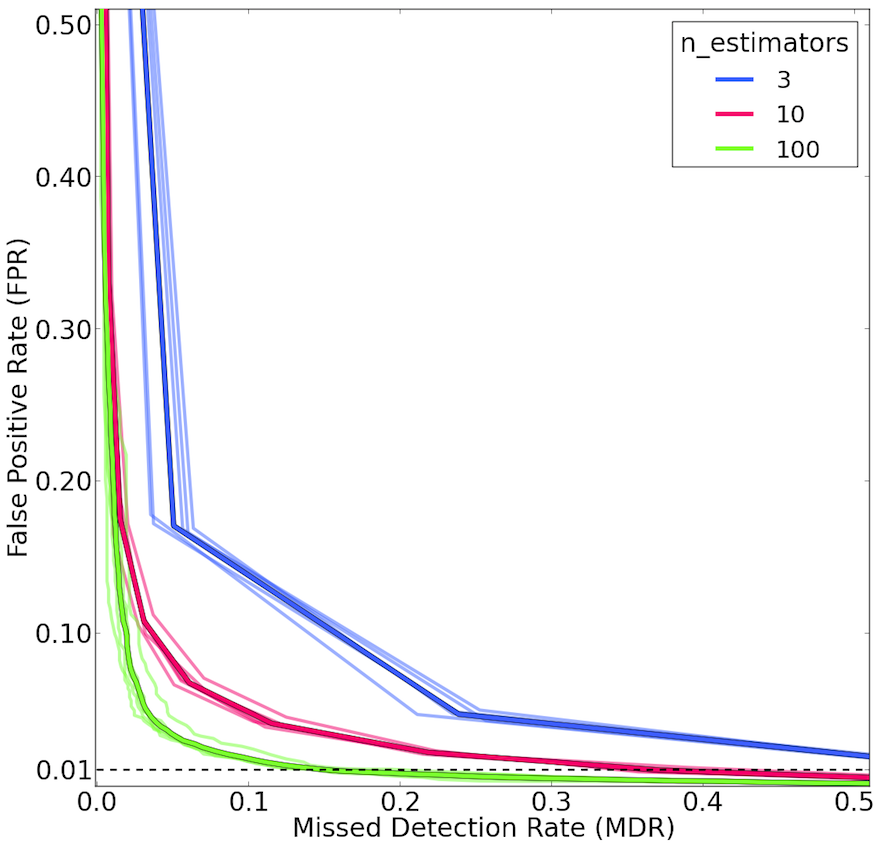}
   \caption{An example of the cross
   validation process for a RF with \texttt{max\_features}=25 and 
\texttt{min\_samples\_leaf}=1.}
   \label{fig:cv} 
\end{figure}

When calculating
the FoM to compare the relative performance of models, it is important
that the measurement is made on data that the model has not inspected
during the training phase, otherwise we risk measuring the performance
on data that the model has overfit and report an FoM that we cannot
expect to achieve on out of sample data.  To mitigate this effect we
split the data we designated for training in
Section~\ref{sec:trainingSet} into 5 subsets or \textit{folds} with
equal numbers of training examples.  We then train each model on 4 of
these folds and use the fifth as a validation set to measure the
performance.  The model is then retrained on 4 folds but a different
fold is held out.  In total the model is trained 5 times with each fold
being held out once.  We then average the results for the 5-folds and
choose the model that results in the best average FoM.  A second
advantage is that for relatively small data sets where the composition
of the validation set may not be representative of the entire
population, by evaluating the performance on each fold in turn and then
averaging, we achieve a better estimate of the actual performance on the
entire data set.\\ 
\indent In our case all 3 classifiers output a
prediction or \textit{hypothesis} for each example.  These hypotheses
can be thought of as the probability a given example
has of belonging to the class of real images, taking on values in the
range 0-1.  A classifier predicts detections with hypotheses close to 1 are
highly likely real transients, while those close to 0 are bogus.  In
Fig.~\ref{fig:nnHypo} we plot the distribution of hypothesis values for
a RF with \texttt{n\_estimators}=100, \texttt{max\_features}=25 and 
\texttt{min\_samples\_leaf}=1 trained on 4 folds of the training set.  The distribution
plotted shows the hypotheses for the held-out fifth fold.  To assign a
label of real or bogus we must define a decision boundary; a hypothesis
value above which the classifier labels detections as real, otherwise detections
are labelled bogus.  If the classifier has learnt a useful model it
should output detections labelled as bogus with a hypothesis below the
decision boundary and those labelled as real above the decision boundary
for the prediction to be correct.   Bogus detections with predictions above
the decision boundary are False Positives and real detections with
hypotheses below the decision boundary are Missed Detections. For our
FoM the decision boundary is selected as the hypothesis value above
which only 1\% of the bogus detections lie (dashed line in
Fig.~\ref{fig:nnHypo}).  The FoM is the fraction of the detections labelled
as real that lie below this choice of decision boundary.  During 5-fold
cross validation a hypothesis distribution is generated by predicting
hypotheses for the detections in each of the held-out folds.\\
\indent In Fig.~\ref{fig:cv} we show an example of the 5-fold cross
validation process for an RF with \texttt{max\_features}=25 and 
\texttt{min\_samples\_leaf}=1.  In
this example we vary the number of decision trees, \texttt{n\_estimators} and plot a
Receiver Operator Characteristic (ROC) curve for each model.  ROC curves
are produced by varying the decision boundary at which we assign a
prediction to a label of real or bogus and calculate the FPR and MDR
that decision boundary produces for the validation set.  From the
example in Fig.~\ref{fig:cv} we see that selecting a value of
100 for \texttt{n\_estimators} produces the best FoM of $\sim$0.167, this means that
an FPR=1\% produces a MDR of 16.7\%.  We also include 5\% and 10\% FPR
levels for reference.  We repeated this process for various sizes of
hidden layer.  We also show an example of measuring the FoM on a data
set containing a significant proportion of the training data, labelled
as overfit in Fig.~\ref{fig:cv}.\\ 
\indent By replicating this
process for both ANNs and SVMs we were able
to select the optimal set of hyperparameters for each algorithm.  In the second column of
Table~\ref{tab:modelResults} we show the optimal hyperparamters selected
for each algorithm by cross validation.  By using the validation sets to
select the hyperparameters, there is a danger that the hyperparameters
will in effect have been fit to these sets.  As a result, the FoM we
measure on the validation sets is not an unbiased measurement of the
performance we would expect to achieve on data not included in the
training folds.  We deal with this in the next section.

\subsubsection{Testing} 
\label{sec:testing} 
\begin{figure}
   \textbf{a)}\\
    \includegraphics[width=84mm]{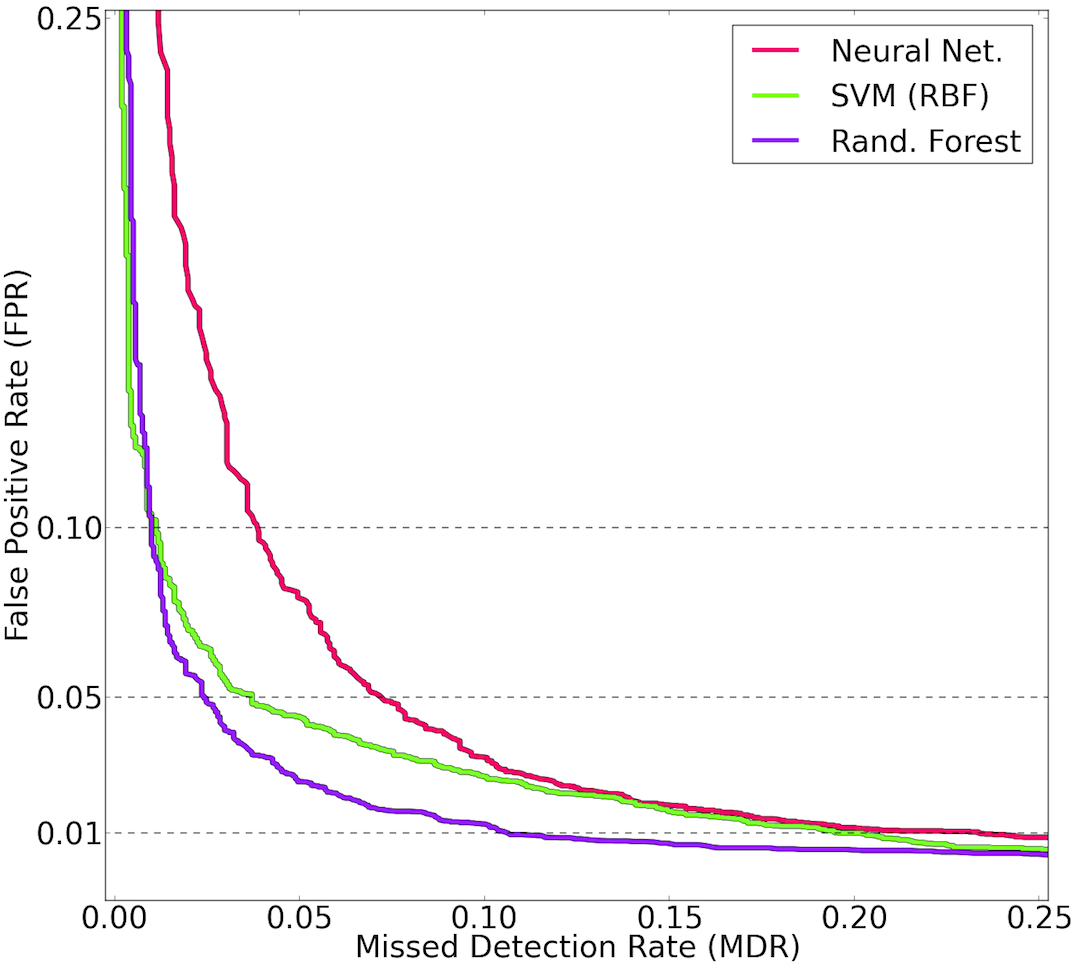}\\
    \textbf{b)}\\
   \includegraphics[width=84mm]{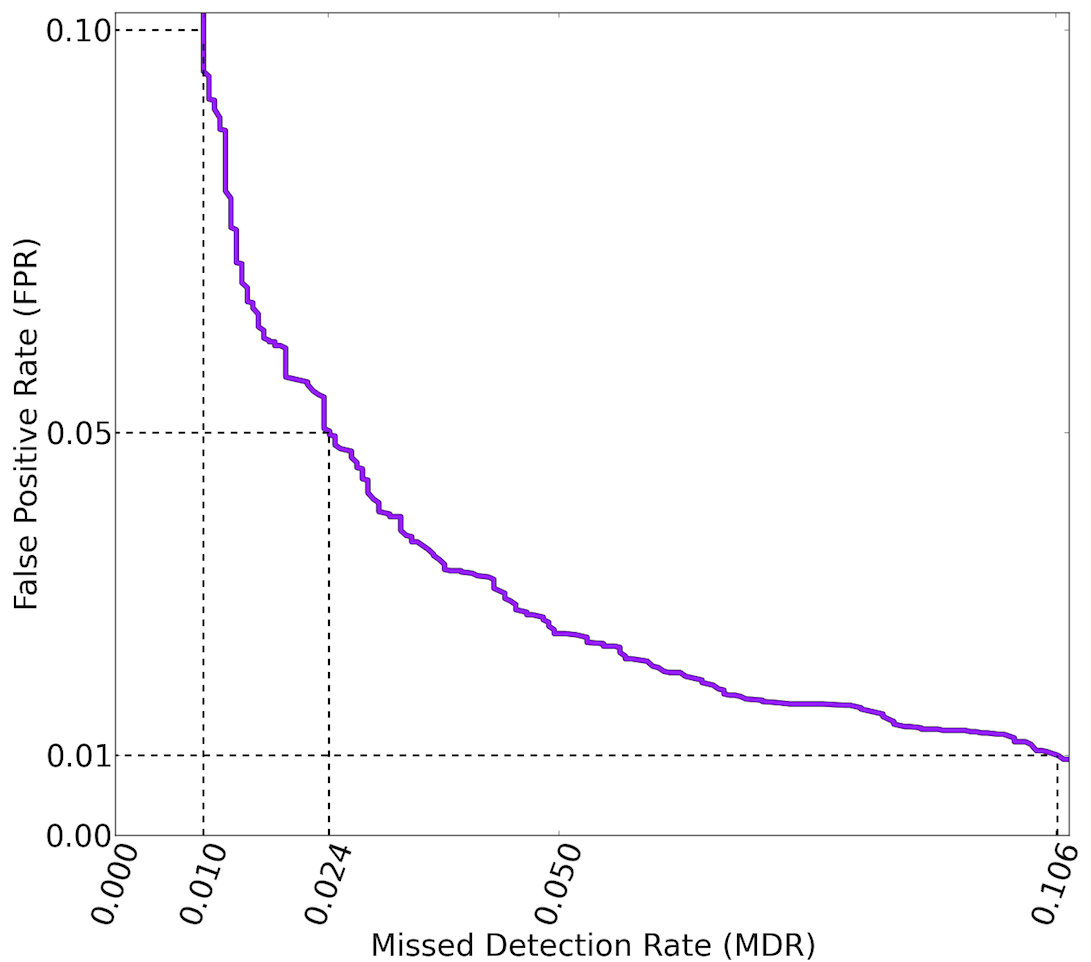} 
    \caption{a) Comparison of the best
    models for various learning algorithms applied to the held out test
    set.  b) Detail of ROC curve of the best performing classifier, the
  Random Forest shown in a).  At a FPR of 1\% the FoM shows that in
  practice we expect to operate at a MDR of 10.6\%.} 
    \label{fig:algoComp} 
\end{figure}

\begin{figure}
   \includegraphics[width=84mm]{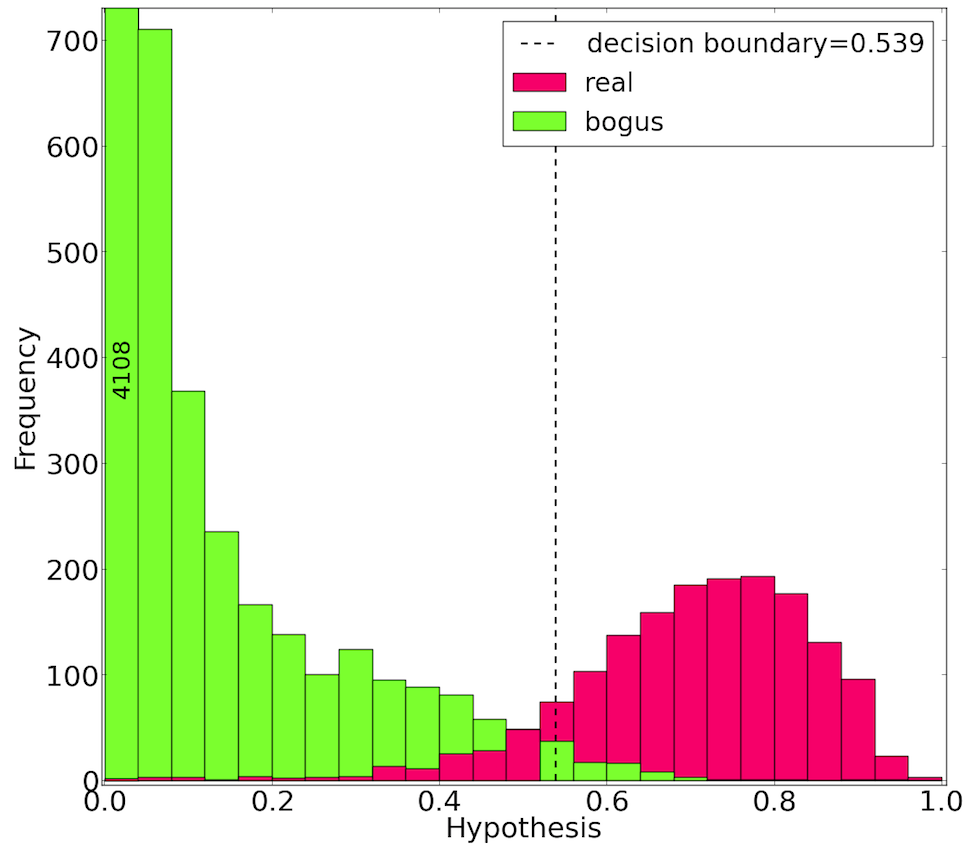} 
   \caption{Hypothesis
   distribution for the optimal  Random Forest classifier applied to the
   test set.} 
   \label{fig:RFHypo} 
\end{figure}

Having selected the optimal
model for each of the algorithms we retrain these models with the entire
training set.  This allows the models to learn from more examples.  To
measure how well we expect the models selected by cross validation in
the last section to perform on unseen data we measure the FoM on the
test set, the 25\% of the data we held back from both training and
validation.  This provides an unbiased estimate of the performance.  In
Table~\ref{tab:modelResults} we show the FoM measured on the test set. 
Fig.~\ref{fig:algoComp}a shows the ROC curve for each model in
Table~\ref{tab:modelResults}.  We find that the RF is the best
classifier with a FoM of 0.106.\\ 
\indent Fig.~\ref{fig:algoComp}b shows a
close-up of the measured FoM for the RF classifier, where the measured
FoM is shown along with the performance we would expect to achieve if we
were to allow 5\% or 10\% of the bogus detections through to human scanners.
 For example, allowing the FPR to slip to 5\% increases the completeness
to 97.6\%.  We also plot the hypothesis distribution for the detections in
the test set in Fig.~\ref{fig:RFHypo}.\\ 
\indent The FoM shown in
Fig.~\ref{fig:algoComp}b is the single best classifier we find in our
analysis.  Using this classifier on a data stream of nightly
observations from PS1, we would expect that 99\% of the detections promoted
to humans would be of real astrophysical transients while 10.6\% of the
real detections would be rejected by the classifier. \citet{Brink13} report
a MDR of 7.7\% for their system.  As a next step it is useful to
investigate the detections for which the classifier produces incorrect
predictions to see if there are systematic errors that the classifier
makes or if it is making correct predictions for detections that have been
labelled incorrectly during the construction of the training set.

\begin{table*} 
\begin{minipage}{145mm} 
\caption{Comparison of learning
algorithms.} 
\begin{tabular}{lcccc} \hline Classifier & Model Parameters & Threshold & FoM\\
\hline \hline 
 Artificial Neural Network  & $s_2$=200, $\lambda=5$ & 0.547 & 0.233\\ 
 Support Vector Machine (RBF)  & \texttt{C}=3, \texttt{gamma}=0.01 & 0.788 & 0.196\\ 
 \textbf{Random Forest} & \textbf{\texttt{n\_estimators}=1000, \texttt{max\_features}=25, \texttt{min\_samples\_leaf}=1} & \textbf{0.539} & \textbf{0.106}\\
\hline
 \end{tabular} 
 \label{tab:modelResults}
\end{minipage} 
\end{table*}

\section{Further Analysis} 
\label{sec:furtherAnalysis} 
In this section
we attempt to get a better sense of how we expect the classifier to
perform in practice by characterising its performance under various
conditions.  We aim to identify trends in the kinds of detections for which
the classifier makes incorrect predictions and investigate the effect
that providing the classifier with incorrectly labelled training and
test sets has on the measured FoM.  However, we begin this section by
looking at methods to boost performance by combining classifiers.

\subsection{Combining Classifiers} 
\label{sec:combining} 
\begin{figure}
   \includegraphics[width=84mm]{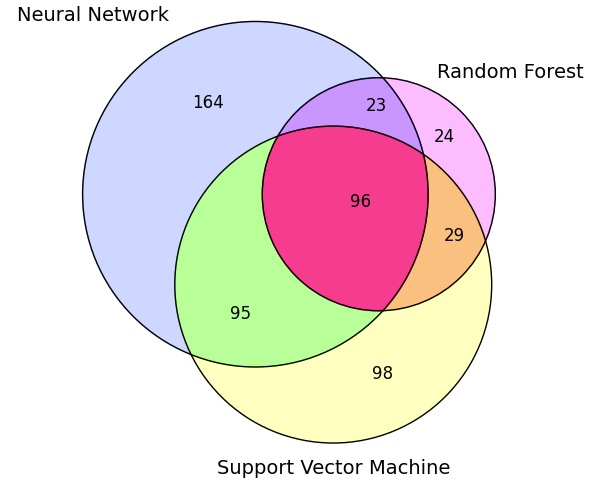} 
   \caption{Venn diagram showing the
   relationship between the Missed Detections for each classifier. 
   There are 1619 positive examples in the test set.} 
   \label{fig:Venn}
\end{figure}

As a last step
toward boosting performance we investigated a selection of methods to
combine the RF, SVM and ANN from
Table~\ref{tab:modelResults}.  The predictions of the 3 methods are
correlated; a candidate highly ranked by the RF is likely to also be
highly ranked by the other 2 classifiers, but there are still detections of
real transients that are discarded by only one of the classifiers.  
From Fig.~\ref{fig:Venn} there are 24 detections labelled as real that only
the RF wrongly rejects, it is these examples that we hope to recover by
combining classifiers.\\ 
\indent We tried only a few of the simplest
combination strategies.  First we simply classified a detection based on
the majority vote of the 3 classifiers.  Second we assigned each detection a
hypothesis that was the mean of the hypothesis values output by each
classifier.  This produced a new distribution of mean hypotheses, where
we again selected the decision boundary to produce the FoM.  Finally we
trained a SVM using the 3 hypotheses for each detection as the features
representing that detection.  In the end none of these methods outperformed
the RF classifier, though the performance was comparable (see
Table~\ref{tab:avgTable}).\\ 
\indent This result is unsurprising given
that the classifiers are highly correlated and there is no guarantee
that these methods will outperform the best individual classifier
\citep{Fumera05}.  The RF is in itself an ensemble of classifiers (the
individual decision trees) and may already incorporate much of the gain
in performance we can expect from these simple methods.

\begin{table}
\centering
    \caption{Results of combining classifiers.} \begin{tabular}{lccccc}
    \hline Method && FPR && MDR\\
    \hline \hline
    Majority Vote && 0.02 && 0.06\\
    Mean Hypotheses && 0.01 && 0.12\\
    Hypotheses as Features && 0.01 && 0.12\\ 
    \end{tabular}
    \label{tab:avgTable}
\end{table}

\subsection{Relative Feature Importance} 
\label{sec:importance} 
\begin{figure}
   \includegraphics[width=84mm]{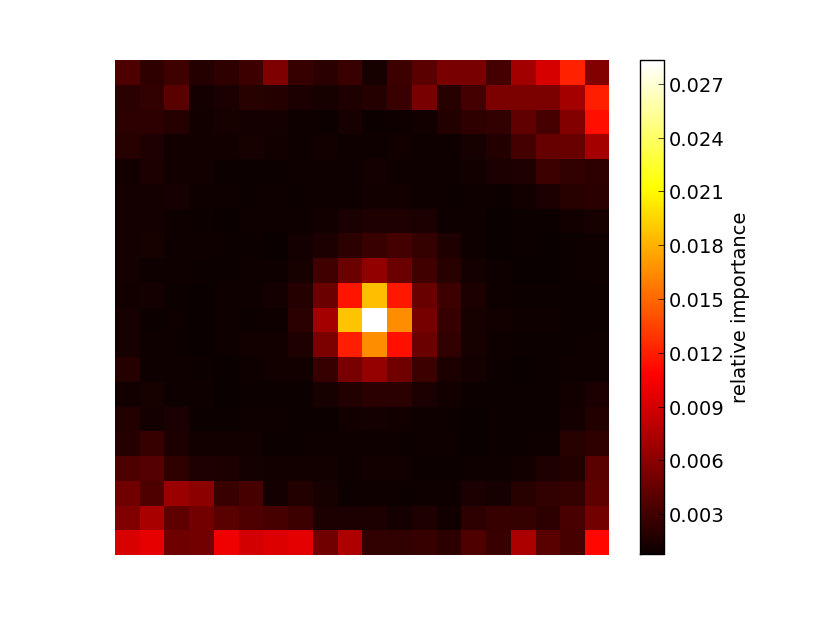} 
   \caption{The relative
   importance of each pixel to the classification for the Random Forest.
    The contributions of each feature are normalised such that they sum
   to 1.} 
   \label{fig:relImportance} 
\end{figure}

Random
Forests provide a built-in method to estimate the relative importance of
each feature to the classification \citep{Breiman01}.  By inspecting the
`depth' at which each feature is used as a decision node we can estimate
the relative importance of that feature, as those features used closer
to the top of the tree will contribute to the prediction of a larger
fraction of the training examples.  The fraction of samples for which we
expect a feature to contribute to the classification can be used to
gauge its relative importance.\\ 
\indent Fig.~\ref{fig:relImportance}
shows the relative importance of each pixel determined from the training
set.  The relative importance metric is normalised such that it sums to
1.  The most important features have the highest values and as would be
expected are located in the centre of the image.  The pixels on the
edges of the images are thought to be important for identifying many of
the bogus examples, where the object is not centred in the substamp and
often lies at the edge.  For reference if features were equally
important they would each have a relative importance of $1/400 =
0.0025$.\\ 
\indent Fig.~\ref{fig:relImportance} may suggest some
redundancy in the features bounding the central pixels.  It is expected
that omitting these features would have little effect on the performance
of our classifier as RFs are thought to be unaffected by the inclusion
of noise variables in the feature vector \citep{Biau10}.  In contrast
\citet{Brink13} find that the MDR for their RF classifier improves by
$\sim$4\% by omitting noisy features using a backward feature selection
method.  The effect of feature selection is an interesting area for
future work and attempts at optimisation.

\subsection{Label Contamination} 
\label{sec:contamination} 
\begin{figure}
   \includegraphics[width=84mm]{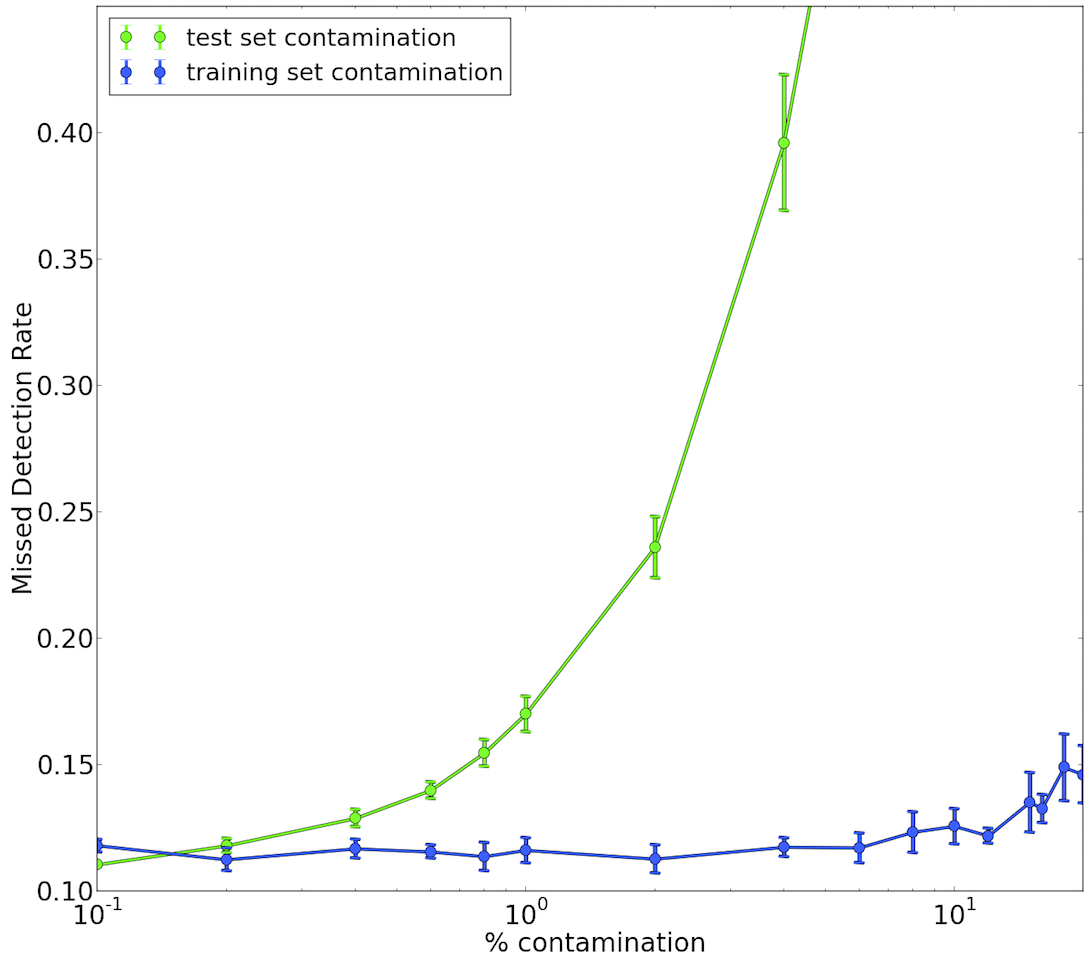} 
   \caption{The effect of randomly
   flipping labels.  As we increase the fraction of the images for which
   we flip the label i.e. as we introduce more label contamination, the
   performance of the classifier trained on the contaminated training
   set and measured on the untouched test set (blue line) decreases as
   expected.  Introducing contamination into the test set has
   a much more pronounced effect on the measured performance even at low
   fractions (green line).} 
   \label{fig:contamination}
\end{figure}

We took care
to eliminate label contamination in Section~\ref{sec:trainingSet}, by
visually checking and manually labelling each training example. 
Nonetheless we expect that there remain some examples with incorrect
labels.  In this section we employ similar methods to those in
\citet{Brink13} to investigate the effect that label contamination has
on our ability to train and test the optimal RF model.\\ 
\indent First
we investigate the effect of adding label contamination to the training
set.  We add contamination by randomly selecting a subset of the detections
from the training set and flipping their labels.  Those labelled as real
are now labelled as bogus and vice versa.  In
Fig.~\ref{fig:contamination} we plot the effect of randomly flipping
labels in the training set while leaving the original labels in the test
set untouched.  The measured MDR appears fairly unaffected up to around
6\% contamination.  The approach of \citet{Brink13} is robust to around
10\% suggesting our method may be more susceptible to incorrectly
labelled training data.\\ 
\indent Next we flip labels in the test set,
while using the original training set labels as they are.  Given that
the RF has been trained with correctly labelled data, for the most part
we expect it to provide the correct labels for the images in the test
set.  However, the flipped labels affect our ability to accurately
measure the FoM. Although the classifier makes sensible predictions,
when we compare these predictions to the flipped labels the otherwise
correct predictions are now evaluated as False Positives or Missed
Detections.  Fig.~\ref{fig:contamination} shows how the FoM is affected
as we increase the fraction of flipped labels, we see that even at low
proportions labelling noise in the test set can have a significant
effect.

\subsection{Classification as a Function of Signal-to-Noise}
\label{sec:signal} 
\begin{figure}
   \includegraphics[width=84mm]{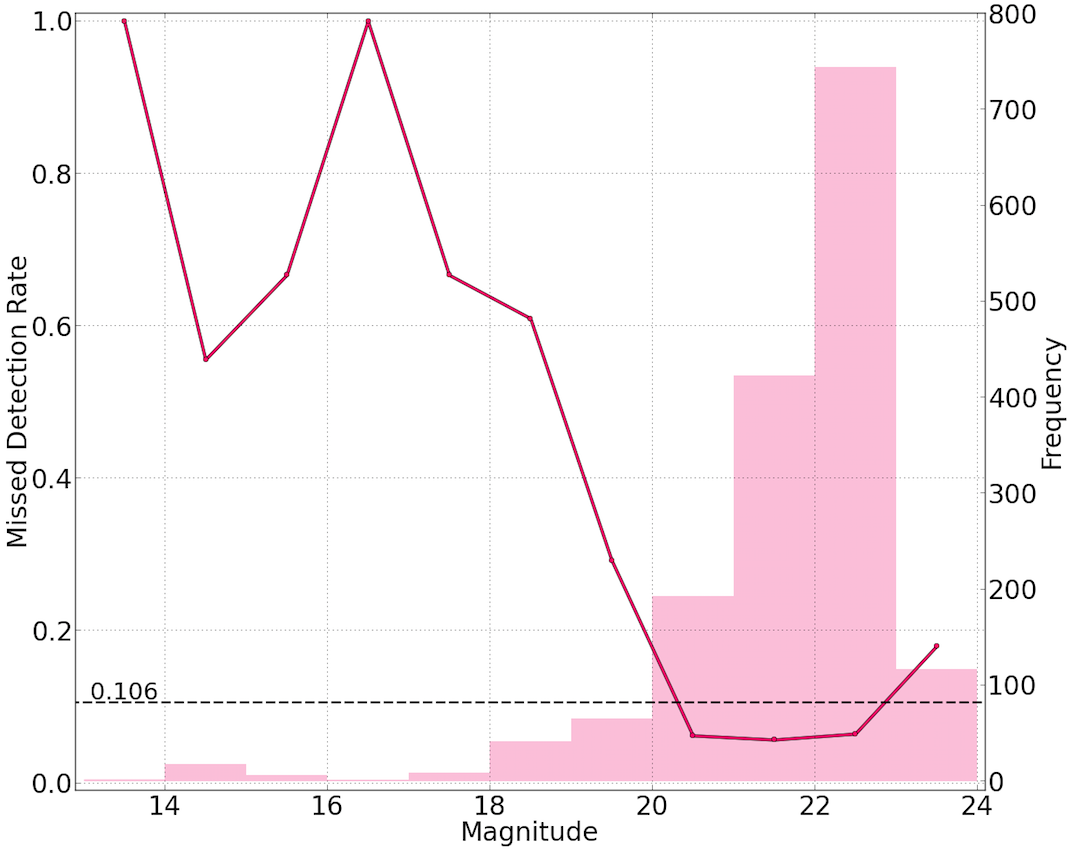} 
   \caption{Histogram of magnitudes for the
   test set examples labelled as real.  We also show the MDR as a
   function of S/N which increases dramatically for sources brighter
   than magnitude 20.} 
   \label{fig:SNplot} 
   \end{figure}

To investigate the classifier performance as a
function of Signal-to-Noise (S/N), we also follow a similar analysis to
\citet{Brink13}.  We plot the distribution of magnitudes for each
example in the test set labelled as real in Fig.~\ref{fig:SNplot}.  We
divide the examples into 11 bins, each spanning 1 magnitude in the range
13 to 24 mag.  We then use the classifier to make a prediction for the
examples in each bin and calculate the fraction of examples classified
as bogus which we take as an estimate of the classifier performance for
objects at that level of S/N. For objects with magnitudes $\gtrsim$20 
there is a $\sim$6\% chance of missing real detections. 
Counterintuitively the detection performance deteriorates for higher S/N
objects.  The number of examples of these cases are low as typically
these objects result in artefacts from saturation and subsequent masking
or unclean subtractions.  However, this can also be understood as an
effect of our feature representation, where we are learning
classifications based on the relative intensity of pixels across the
substamp.  The
tendency to misclassify such detections could stem from a combination of
the large relative intensity differences between pixels in these substamps
that often characterise artefacts and the low numbers of high S/N images
of real transients.  This explanation is further supported by both the
ANN and SVM, which also misclassify these objects, suggesting the issue
is with the data and not a consequence of the realisation of the RF.  In
the next section we try to identify any relationships in the missed
detections.

\subsection{Missed Detections} 
\label{sec:missedDetections} 
\begin{figure*} 
\vspace{20pt}
\includegraphics[width=168mm]{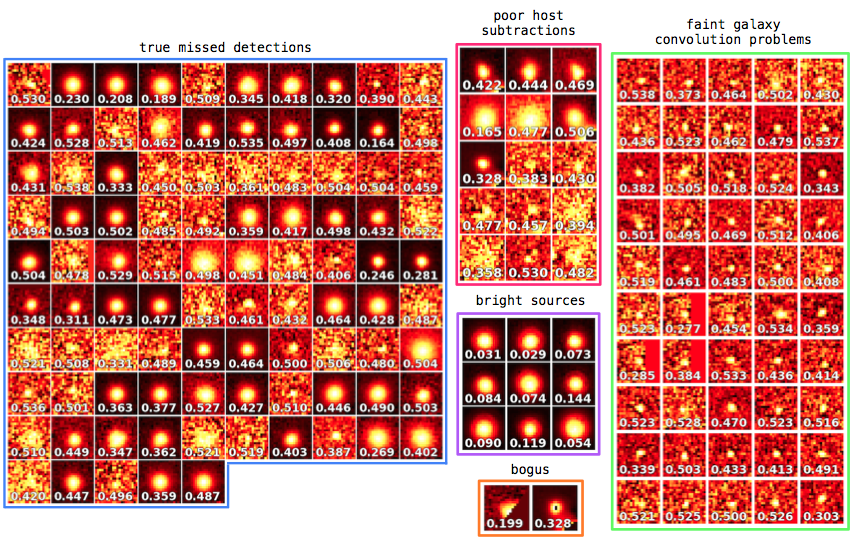} 
\caption{The 172 detections
labelled as real but classified as bogus by the RF.  Detections are
grouped according to the discussion in
Section~\ref{sec:missedDetections}.  The hypotheses for the detections
are shown as the inset numbers.}
\label{fig:missedDetections} 
\end{figure*}

\begin{figure}
\includegraphics[width=84mm]{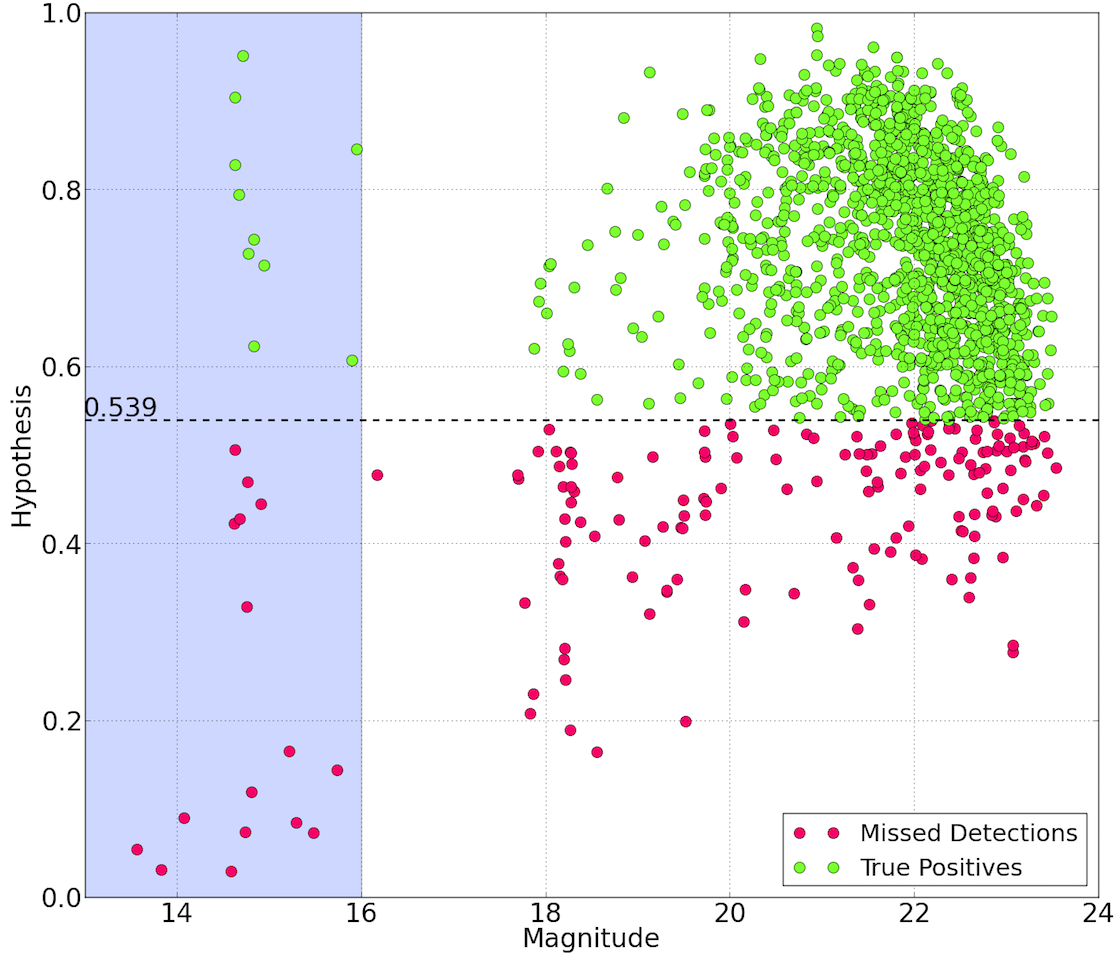} 
\caption{Plot showing the hypotheses of the
real examples in the test set against the magnitude measured by IPP. 
The shaded region shows the magnitude cut above which we cannot be
certain the nightly stacks do not contain saturated images.}
\label{fig:predVsMag} 
\end{figure}

\begin{figure}
\includegraphics[width=84mm]{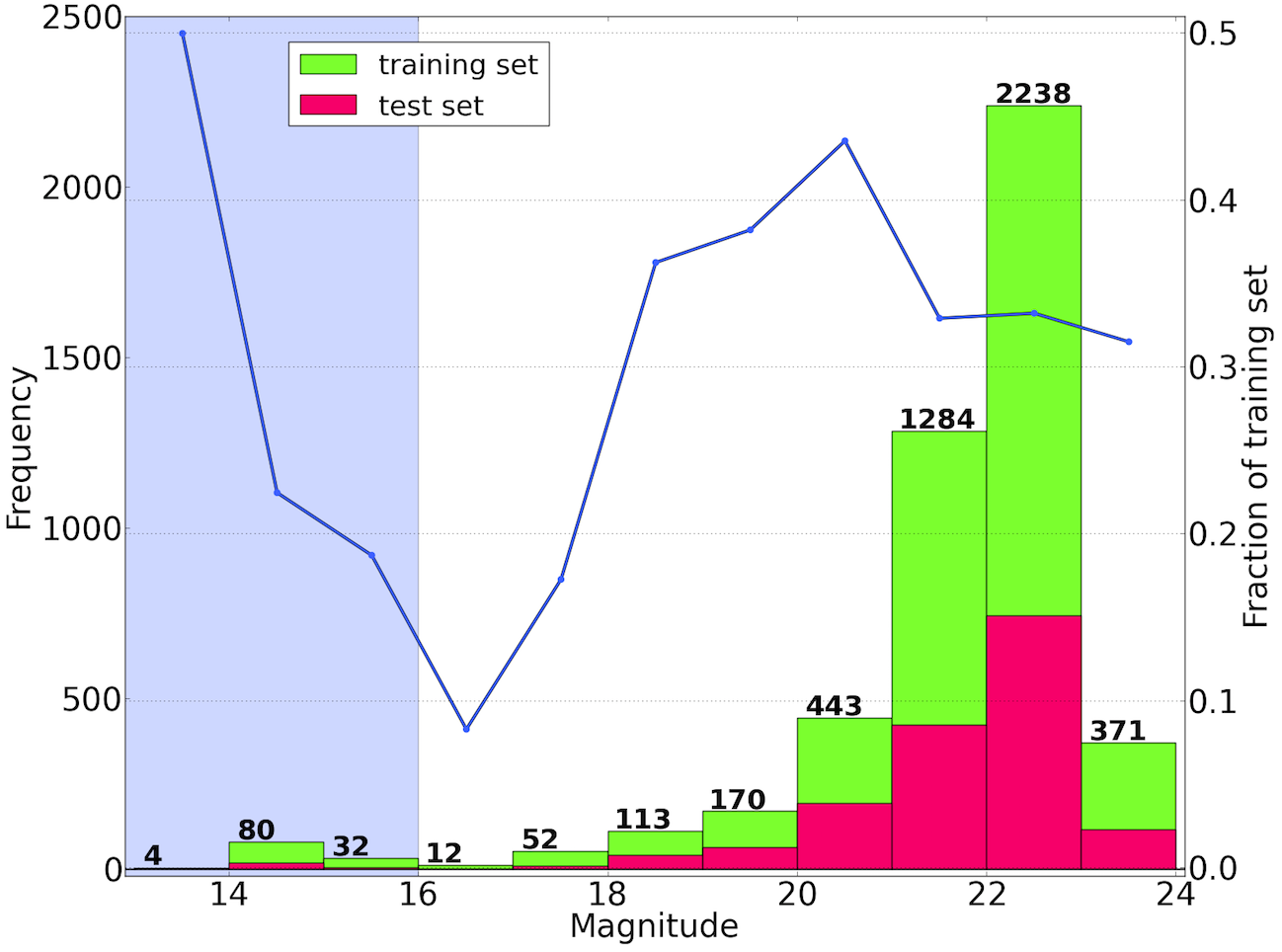} 
\caption{The magnitude distributions of the real
examples in the training and test sets.  The numbers on each bin show
the total number of images in the training set.  We also show the
relative numbers of test set examples in each bin (blue line).  The
shaded region again shows the magnitude cut defined in
Section~\ref{sec:missedDetections}} 
\label{fig:magDist} 
\end{figure}

We inspected
the 172 missed detections (see Fig.~\ref{fig:missedDetections}) looking
for similarities that may explain why they were rejected.  We found that these
missed detections are associated with 112 individual transients.    Although we
took care to limit label contamination during the construction of the
training set, we identified some examples of obvious bogus detections mislabelled as
real that account for a  small fraction ($\sim$1\%) of the missed
detections.\\
\indent We also find about 29\% of the missed detections appear to be a result of
faint galaxy convolution problems (see Section~\ref{sec:artefacts}).  These artefacts are difficult
to identify by eye and as a result have been incorrectly labelled as
real detections significantly contributing to the label contamination of the test set.\\
\indent In Section
~\ref{sec:signal} we discussed the high MDRs for bright sources.  In
Fig.~\ref{fig:predVsMag} we plot the hypothesis values for all detections
included in the histogram of Fig.~\ref{fig:SNplot} (i.e. all test set
detections that have been visually classified as real) against their
magnitude reported by IPP.  A feature of the plot that stands out is the
cluster of sources with magnitudes brighter than 16 and hypotheses less
than 0.2.  \citet{Magnier13} report that for the PS1 3$\pi$ survey,
saturation occurs at $\sim$13.5 for g$_{P1}$, r$_{P1}$, i$_{P1}$,
$\sim$13.0 for z$_{P1}$ and $\sim$12.0 in y$_{P1}$.  We were concerned
that these sources could be saturated, however to conclusively determine
this the individual images that are combined to make a nightly stack
would need to be examined.  Instead we scaled the magnitudes reported by
\citet{Magnier13} for PS1 3$\pi$ exposures by the exposure times for the
individual images that make up a nightly stack and set a magnitude limit
of 16 mag.  Objects brighter than this limit may have saturated cores in some
exposures and cannot safely be labelled as real.  Some of these sources
on close inspection also show signs of the unclean subtractions we
highlighted in Section~\ref{sec:artefacts}.\\
\indent The detections brighter than 16 mag in Fig.~\ref{fig:predVsMag} with
hypotheses above 0.2 are all associated with a single confirmed
supernova (SN), SN 2014bc (PS1-14xz)\citep{SmarttAtel14}.  SN 2014bc is a
nearby (7.6 Mpc) Type-IIP located in the bright host galaxy NGC4258
(Messier 106).  The transient lies close to the core of the host and as
a consequence the host has been poorly subtracted in the same location
in all the substamps.  Detections of this object appear in both the
training and test set and although we ensured detections from the same night
must appear in the same set, the slowly evolving plateau has resulted in
detections with similar S/N and the same pattern of poor subtraction
appearing in both.  It is therefore to be suspected that test set detections
associated with this SN would have been rejected along with the other
sources brighter than 16 mags had similar detections not been included in
the training set.  This raises the issue of potentially missing the
brightest transients which are often of interest and the cheapest to
classify spectroscopically, we return to this in
Section~\ref{sec:summary}.\\
\indent The high MDRs in the
magnitude range 16-20 still remain unexplained.  To address this in
Fig.~\ref{fig:magDist} we plot the number of examples of real transients
in each of the magnitude bins used in Section~\ref{sec:signal} for both
the training and test sets.  The plot clearly shows the deficit in
training examples at magnitudes brighter than 20 and lead us to conclude
that we lack enough training examples of high S/N transients to allow
the classifier to learn a model that generalises well in this regime. 
In Fig.~\ref{fig:magDist} we overlay the relative size of the test set
compared with the training set in each bin.  We selected the test set by
randomly sampling 25\% of the data available for training.  The small
fractions of test examples available between 16 and 18 mags combined
with the low numbers in the range 16-20 mags severely impact our ability
to accurately measure the MDR in this range.\\
\indent Aside from the issues associated with high S/N, there are a few other SNe with detections that 
show similar host galaxy subtraction problems to SN 2014bc.  Some of these are true bogus detections
which we show in Fig.~\ref{fig:missedDetections}.
Approximately 9\% of the missed detections are bogus detections
around poor host subtractions.  We include detections of SN 2014bc
with this group in Fig.\ref{fig:missedDetections} though these
detections around 15th magnitude could
equally have been included with the bright sources.\\
\indent Among the missed detections we also found substamps
where entire rows or columns along an edge of a substamp had been masked. 
In the second panel of Fig.~\ref{fig:realVectors} we show an example
where the bottom 2 rows of pixels have been masked.  These are
examples of the sky cell duplicates we describe in
Section~\ref{sec:artefacts}.  We were concerned the classifier was rejecting
these detections based on the masking.  To see if this was
the case we identified all the examples of sky cell overlap among the
real test set detections, and found 20.  As we ensured that detections from the
same night must be in the same set (training or test set), the
equivalent full 20$\times$20 pixel substamps were also in the test set.  We
compared the performance on the full pixel substamp with that of the
partially masked substamp and found that there is only 1 case where the
masked substamp was rejected while the full pixel substamp was kept.  In this
instance a significant proportion of the substamp was masked (7 columns)
with the edge lying close to the PSF.  The majority of the remaining
substamp pairs were both assigned the same classification.  There are
however 6 pairs where the masked substamp was correctly classified as real,
but the full pixel substamp was rejected, showing the classifier does not
tend to reject detections with sky cell masking simply due to the masked
regions.\\ 
\indent The reason for rejecting one detection from the pair over
the other is unclear as both substamps are constructed from the same data. 
The 6 pairs for which the full pixel substamp was labelled bogus, but the
masked substamp was labelled real are all associated with a single
transient and may not apply to other sky cell pairs.  For
these substamp pairs we found that the centroids always differed by 1 pixel
and were offset in the same direction.  We tried shifting the centre of
the stamps to the same pixel, but found that this had little impact on
the hypothesis.  In all cases the flux-conserving warping results in
equivalent pixels containing different counts, though the difference is
typically small $\lesssim$10\%.  Given the small number of cases where
the detections of a sky cell pair are assigned to different classes (7 in
total) and that these detections are associated with only two transients (6
associated with a single transient where the full pixel substamp is
rejected and 1 associated with a different transient where the masked
substamp is rejected), it is difficult to explain this behaviour, though
one explanation may be the small differences in pixel intensity values
perhaps combined with the different centroids.\\
\indent The nuances of difference imaging make it difficult to
determine the ground truth label for each detection.  Humans 
often require additional information beyond that contained in the single difference image e.g. position relative to the host, or the number of bad/good pixels visible in the input image.  The investigations above suggest that the 
classifier is identifying subtle relationships and correctly identifying that many of the `missed detections' are dubiously labelled
 as real.  We estimate that 45\% ($\sim$5\% bright sources; $\sim$29\% convolution problems; $\sim$9\% 
poor host subtractions; $\sim$1\% obvious mislabelled artefacts) or about 77 of the missed detections 
are not of high enough quality to be confidently labelled as real
detections.  Therefore the RF classifier is not strictly getting them
wrong.  The high proportions of these cases among the
missed detections does not hold true for the entire sample of real
detections in the test set, where for example faint galaxy convolution problems
are crudely estimated to account for no more than 7\%.  Removing such detections from our test set results 
in an MDR around 6.2\%. The MDR of our classifier is therefore in the
range 6.2 - 10.6\%  for a FPR of 1\% but most likely toward the lower
end of this range.  The remaining 95 detections are true missed detections and
appear to be mislabelled by the classifier due to high S/N as
discussed above, poor seeing conditions and very low S/N detections near the
detection limit.

\subsection{Medium Deep Confirmed Supernovae} 
\label{sec:lightcurve} 
\begin{figure}
   \includegraphics[width=84mm]{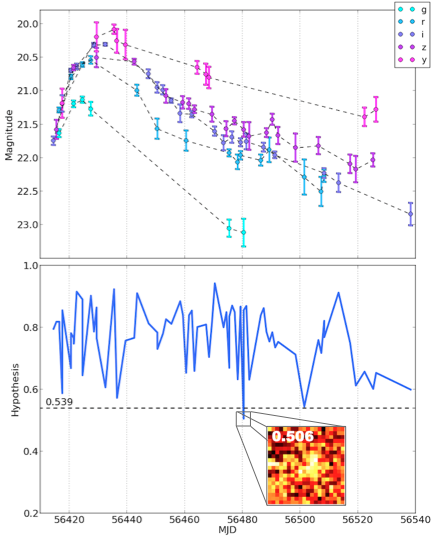}
   \caption{(top panel) PS1 lightcurve of the Type Ib Supernova
   PS1-13avb.  (bottom panel) The hypothesis for each epoch.  The dashed line shows the decision boundary (0.539) below which the
   classifier predicts an image as bogus.  (inset) The only missed
   detection for this SN which shows low S/N.} 
   \label{fig:lightcurve}
\end{figure}

In
order to demonstrate how we might expect the classifier to perform on a
live data stream we first use the classifier to make predictions for the
supernova PS1-13avb for which we held out all associated detections from
both the test and training set.  This object has been spectroscopically
classified as a Type Ib SN and has a well sampled lightcurve from
about -18 days pre-maximum to around 106 days post maximum, including
exposures in all 5 filters ranging in magnitude from around 23 to 20
mags (see Fig.~\ref{fig:lightcurve} top panel).  We selected this object
for its high quality lightcurve and magnitude range which represents the
majority of objects discovered in the PS1 MDS.  In the bottom panel of
Fig.~\ref{fig:lightcurve} we show the hypothesis for each epoch of this
target.  The plot shows that the hypothesis is consistently above the
decision boundary of 0.539 (selected in Section~\ref{sec:testing}) with
the exception of the detection from 56480.406 MJD (Modified Julian Date)
which shows the transient at a magnitude of g$_{P1}=23.12\pm0.21$
approaching the detection limit in this filter.  The detection is displayed
as an inset in Fig.~\ref{fig:lightcurve} with its hypothesis of 0.506,
showing the low S/N and deviation from a PSF-like morphology.

\subsection{Early Detection} 
\label{sec:earlydetection} 
One of the major
aims of recent supernova searches has been to try to detect the
transient as soon after explosion as possible in order to trigger rapid
follow-up to spectroscopically study regions of the transients evolution
that remain relatively unexplored (\citet{Gal-Yam14}; \citet{Cao13}). 
To this end we carry out a simple test by using the classifier to make
predictions for the first detections of all 53 classified SNe in our
database.  Again we held these detections out from the training and test
sets.  In Table~\ref{tab:firstDetections} we list the 53 SNe and the
details of the first detections along with the hypothesis for each
detection.  The classifier correctly predicts all detections as real and had it
been running on a live data stream would have promoted all objects to
humans for follow-up.

\section{Summary of Results and Conclusions} 
\label{sec:summary} 
In this work we have constructed a data set of detections from the Pan-STARRS1
Medium Deep Survey.  We used this data set to train a Random Forest
classifier to reject bogus detections of transients before they are
presented to humans as potential targets for follow-up. As the feature
representation of these detections we used the pixel intensity values of a
20$\times$20 pixel substamp centred on the detection.  This choice is
independent of the observing strategy and removes the need for careful
feature design and selection that requires specific domain knowledge. 
The choice of features also make this method applicable to any survey
performing difference imaging and requires no information from either
the template image or nightly stack.  Using the Figure of Merit as defined in
\citet{Brink13} we selected the decision boundary such that objects
classified as real should be 99\% pure, which resulted in a best
estimate of a Missed
Detection Rate of 6.2\% (i.e. 93.8\% complete) and can compete with
previous work in this area.  We further tested the classifier by
applying it to the lightcurve of a Type-Ib supernova and found only one
missed detection out of 74.  The missed detection had low
signal-to-noise.  In addition, to assess the
classifiers performance for early detection, we used the classifier to
make predictions for the first detections of 53 spectroscopically
confirmed SNe in our database and found none would have been rejected.\\
\indent
We discovered our classifier struggles to provide accurate classifications for the brightest sources ($<$19 mags).  Many of these 
are associated with bright variable stars and have ringing patterns due to the kernel  size definition, which leads to labelling difficulties. 
Some are also close to the saturation limit which may cause the algorithms to misidentify real sources as bogus. 
The mathematical problem in detecting bright variable stars in difference images is clearly quite distinct from finding low flux 
and moderate flux level transients in, or near extended galaxies. Furthermore the scientific goal in characterising variability 
of stellar sources is typically based on total flux measurements whereas finding explosive transients requires the resolved
and unresolved galaxies to be subtracted. Our methods are tailored toward the latter, and can certainly not be blindly
applied to uncover complete populations of variable stars or variable AGNs. With a goal of discovering extragalactic transients, 
one is content to ignore stellar variables in a data stream, although we show here that the algorithms can sometimes
misclassify bright and high signal-to-noise explosive transients.\\
\indent We also found the MDR is
consistently higher for sources brighter than 20 mags which we attribute
to the lack of training data in this range.  We would expect that
providing more training examples that are representative of these
objects would reduce the MDR for brighter sources.  In this paper we
have only used a sample of the data from the PS1 MDS,
but we have access to the full database of MDS transients, which could
be used to provide more training data.  In addition we also have data from PS1 3$\pi$ difference imaging which
could also be used to boost training numbers and build a classifier that
could perform real-bogus classification for both surveys. In our analysis 
we have not considered the
case of asteroids as these are typically removed during the construction
of the nightly stacks in the MDS.  Including the 
PS1 3$\pi$ data, where differencing is performed on individual 
exposures, would allow us to test the performance of our method on asteroids.
It may also be more beneficial to apply this approach at the source extraction
stage. By working directly on the pixel data the classifier could
potentially learn which sources to extract and which to discard from a
difference image before any further processing of a potential detection
is performed.\\ 
\indent The
dependence of any machine learning approach to real-bogus classification
on large amounts of training data presents a serious problem for any new
survey.  While many sources of processing artefacts are common across
surveys, differing pixel scales and seeing conditions prevent the use of
a classifier trained on one survey being directly applied to another.
A solution would be to build a training set based on hand
labelled commissioning data and periodically retrain the classifier as
new data become available.  Alternatively an initial classifier
trained on the limited data available early in a survey could be
improved on
by employing online learning, where the classifier is automatically updated as new labelled
data are gathered \citep{Shalev-Shwartz11, Saffari09}.\\
\indent Future work will focus on combining
the remaining PS1 data available into a single training set that will
hopefully address the S/N issue.  Other areas of research could include
the use of semi-supervised feature learning
\citep{Raina07} and deep learning \citep{Coates13} that retain all the
advantages of our current approach at the expense of being more
computationally demanding.  However, the added representational power of
larger ANNs and the possibility of applying the unsupervised features
learnt from one survey to a variety of other surveys could mean this is
a promising domain to explore.\\ 
\indent An efficient real-bogus
classifier is only one step toward rapid discovery and classification of
transients.  With next generation surveys the stream of transients will
need to be prioritised based on scientific goals.  Providing a
contextual classification \citep{Djorgovski12, Bloom12} of the
transients detected would allow researchers to select the most promising
candidates for their research goals and will also be the focus of future
work.

\section*{Acknowledgments} The Pan-STARRS1 Survey has been made possible
through contributions of the Institute for Astronomy, the University of
Hawaii, the Pan-STARRS Project Office, the Max-Planck Society and its
participating institutes, the Max Planck Institute for Astronomy,
Heidelberg and the Max Planck Institute for Extraterrestrial Physics,
Garching, The Johns Hopkins University, Durham University, the
University of Edinburgh, Queen's University Belfast, the
Harvard-Smithsonian Center for Astrophysics, and the Las Cumbres
Observatory Global Telescope Network, Incorporated, the National Central
University of Taiwan, and the National Aeronautics and Space
Administration under Grant No. NNX08AR22G issued through the Planetary
Science Division of the NASA Science Mission Directorate.  The research
leading to these results has received funding from the European Research
Council under the European Union's Seventh Framework Programme
(FP7/2007-2013)/ERC Grant agreement n$^{\rm o}$ [291222]  (PI : S. J.
Smartt) and the RCUK STFC grants ST/I001123/1 and ST/L000709/1.  DEW
acknowledges support from DEL in the form of a postgraduate studentship.

\bibliographystyle{mn2e} 
\bibliography{ps1_real_bogus}

\begin{thebibliography}{46}
\expandafter\ifx\csname natexlab\endcsname\relax\def\natexlab#1{#1}\fi

\bibitem[{{Bailey} {et~al}\mbox{.}(2007){Bailey}, {Aragon}, {Romano}, {Thomas},
  {Weaver}, \& {Wong}}]{Bailey07}
{Bailey} S., {Aragon} C., {Romano} R., {Thomas} R.~C., {Weaver} B.~A., {Wong}
  D., 2007, ApJ, 665, 1246

\bibitem[{{Baltay} {et~al}\mbox{.}(2013){Baltay}, {Rabinowitz}, {Hadjiyska},
  {Walker}, {Nugent}, {Coppi}, {Ellman}, {Feindt}, {McKinnon}, {Horowitz}, \&
  {Effron}}]{Baltay13}
{Baltay} C. {et~al.}, 2013, PASP, 125, 683

\bibitem[{{Berger} {et~al}\mbox{.}(2012){Berger}, {Chornock}, {Lunnan},
  {Foley}, {Czekala}, {Rest}, {Leibler}, {Soderberg}, {Roth}, {Narayan},
  {Huber}, {Milisavljevic}, {Sanders}, {Drout}, {Margutti}, {Kirshner},
  {Marion}, {Challis}, {Riess}, {Smartt}, {Burgett}, {Hodapp}, {Heasley},
  {Kaiser}, {Kudritzki}, {Magnier}, {McCrum}, {Price}, {Smith}, {Tonry}, \&
  {Wainscoat}}]{Berger12}
{Berger} E. {et~al.}, 2012, ApJL, 755, L29

\bibitem[{{Bertin}(2001)}]{Bertin01EyE}
{Bertin} E., 2001, in Mining the Sky, {Banday} A.~J., {Zaroubi} S.,
  {Bartelmann} M., eds., p. 353

\bibitem[{{Biau}(2010)}]{Biau10}
{Biau} G., 2010, ArXiv e-prints (1005.0208)

\bibitem[{{Bloom} {et~al}\mbox{.}(2012){Bloom}, {Richards}, {Nugent}, {Quimby},
  {Kasliwal}, {Starr}, {Poznanski}, {Ofek}, {Cenko}, {Butler}, {Kulkarni},
  {Gal-Yam}, \& {Law}}]{Bloom12}
{Bloom} J.~S. {et~al.}, 2012, PASP, 124, 1175

\bibitem[{{Breiman}(2001)}]{Breiman01}
{Breiman} L., 2001, Machine learning, 45, 5

\bibitem[{{Brink} {et~al}\mbox{.}(2013){Brink}, {Richards}, {Poznanski},
  {Bloom}, {Rice}, {Negahban}, \& {Wainwright}}]{Brink13}
{Brink} H., {Richards} J.~W., {Poznanski} D., {Bloom} J.~S., {Rice} J.,
  {Negahban} S., {Wainwright} M., 2013, MNRAS, 435, 1047

\bibitem[{{Cao} {et~al}\mbox{.}(2013){Cao}, {Kasliwal}, {Arcavi}, {Horesh},
  {Hancock}, {Valenti}, {Cenko}, {Kulkarni}, {Gal-Yam}, {Gorbikov}, {Ofek},
  {Sand}, {Yaron}, {Graham}, {Silverman}, {Wheeler}, {Marion}, {Walker},
  {Mazzali}, {Howell}, {Li}, {Kong}, {Bloom}, {Nugent}, {Surace}, {Masci},
  {Carpenter}, {Degenaar}, \& {Gelino}}]{Cao13}
{Cao} Y. {et~al.}, 2013, ApJL, 775, L7

\bibitem[{{Chomiuk} {et~al}\mbox{.}(2011){Chomiuk}, {Chornock}, {Soderberg},
  {Berger}, {Chevalier}, {Foley}, {Huber}, {Narayan}, {Rest}, {Gezari},
  {Kirshner}, {Riess}, {Rodney}, {Smartt}, {Stubbs}, {Tonry}, {Wood-Vasey},
  {Burgett}, {Chambers}, {Czekala}, {Flewelling}, {Forster}, {Kaiser},
  {Kudritzki}, {Magnier}, {Martin}, {Morgan}, {Neill}, {Price}, {Roth},
  {Sanders}, \& {Wainscoat}}]{Chomiuk11}
{Chomiuk} L. {et~al.}, 2011, ApJ, 743, 114

\bibitem[{{Chornock} {et~al}\mbox{.}(2013){Chornock}, {Berger}, {Rest},
  {Milisavljevic}, {Lunnan}, {Foley}, {Soderberg}, {Smartt}, {Burgasser},
  {Challis}, {Chomiuk}, {Czekala}, {Drout}, {Fong}, {Huber}, {Kirshner},
  {Leibler}, {McLeod}, {Marion}, {Narayan}, {Riess}, {Roth}, {Sanders},
  {Scolnic}, {Smith}, {Stubbs}, {Tonry}, {Valenti}, {Burgett}, {Chambers},
  {Hodapp}, {Kaiser}, {Kudritzki}, {Magnier}, \& {Price}}]{Chornock13}
{Chornock} R. {et~al.}, 2013, ApJ, 767, 162

\bibitem[{{Coates} {et~al}\mbox{.}(2013){Coates}, {Huval}, {Wang}, {Wu},
  {Catanzaro}, \& {Ng}}]{Coates13}
{Coates} A., {Huval} B., {Wang} T., {Wu} D., {Catanzaro} B., {Ng} A., 2013, in
  Proceedings of the 30th International Conference on Machine Learning
  (ICML-13), pp. 1337--1345

\bibitem[{{Coates}, {Lee} \& {Ng}(2011){Coates}, {Lee}, \& {Ng}}]{Coates11}
{Coates} A., {Lee} H., {Ng} A.~Y., 2011, in AISTATS 2011, Vol. 1001

\bibitem[{{Cortes} \& {Vapnik}(1995)}]{Cortes95}
{Cortes} C., {Vapnik} V., 1995, Machine learning, 20, 273

\bibitem[{{Djorgovski} {et~al}\mbox{.}(2012){Djorgovski}, {Mahabal}, {Donalek},
  {Graham}, {Drake}, {Moghaddam}, \& {Turmon}}]{Djorgovski12}
{Djorgovski} S.~G., {Mahabal} A.~A., {Donalek} C., {Graham} M.~J., {Drake}
  A.~J., {Moghaddam} B., {Turmon} M., 2012, ArXiv e-prints (1209.1681)

\bibitem[{{Donalek} {et~al}\mbox{.}(2008){Donalek}, {Mahabal}, {Djorgovski},
  {Marney}, {Drake}, {Glikman}, {Graham}, \& {Williams}}]{Donalek08}
{Donalek} C., {Mahabal} A., {Djorgovski} S.~G., {Marney} S., {Drake} A.,
  {Glikman} E., {Graham} M.~J., {Williams} R., 2008, in American Institute of
  Physics Conference Series, Vol. 1082, American Institute of Physics
  Conference Series, {Bailer-Jones} C.~A.~L., ed., pp. 252--256

\bibitem[{{Drake} {et~al}\mbox{.}(2009){Drake}, {Djorgovski}, {Mahabal},
  {Beshore}, {Larson}, {Graham}, {Williams}, {Christensen}, {Catelan},
  {Boattini}, {Gibbs}, {Hill}, \& {Kowalski}}]{Drake09}
{Drake} A.~J. {et~al.}, 2009, ApJ, 696, 870

\bibitem[{{du Buisson} {et~al}\mbox{.}(2014){du Buisson}, {Sivanandam},
  {Bassett}, \& {Smith}}]{duBuisson14}
{du Buisson} L., {Sivanandam} N., {Bassett} B.~A., {Smith} M., 2014, ArXiv
  e-prints (1407.4118)

\bibitem[{{Fumera} \& {Roli}(2005)}]{Fumera05}
{Fumera} G., {Roli} F., 2005, Pattern Analysis and Machine Intelligence, IEEE
  Transactions on, 27, 942

\bibitem[{{Gal-Yam} {et~al}\mbox{.}(2014){Gal-Yam}, {Arcavi}, {Ofek},
  {Ben-Ami}, {Cenko}, {Kasliwal}, {Cao}, {Yaron}, {Tal}, {Silverman}, {Horesh},
  {De Cia}, {Taddia}, {Sollerman}, {Perley}, {Vreeswijk}, {Kulkarni}, {Nugent},
  {Filippenko}, \& {Wheeler}}]{Gal-Yam14}
{Gal-Yam} A. {et~al.}, 2014, Nature, 509, 471

\bibitem[{{Geva} \& {Sitte}(1992)}]{Geva92}
{Geva} S., {Sitte} J., 1992, Neural Networks, IEEE Transactions on, 3, 621

\bibitem[{{Gezari} {et~al}\mbox{.}(2012){Gezari}, {Chornock}, {Rest}, {Huber},
  {Forster}, {Berger}, {Challis}, {Neill}, {Martin}, {Heckman}, {Lawrence},
  {Norman}, {Narayan}, {Foley}, {Marion}, {Scolnic}, {Chomiuk}, {Soderberg},
  {Smith}, {Kirshner}, {Riess}, {Smartt}, {Stubbs}, {Tonry}, {Wood-Vasey},
  {Burgett}, {Chambers}, {Grav}, {Heasley}, {Kaiser}, {Kudritzki}, {Magnier},
  {Morgan}, \& {Price}}]{Gezari12}
{Gezari} S. {et~al.}, 2012, Nature, 485, 217

\bibitem[{{Gezari} {et~al}\mbox{.}(2010){Gezari}, {Rest}, {Huber}, {Narayan},
  {Forster}, {Neill}, {Martin}, {Valenti}, {Smartt}, {Chornock}, {Berger},
  {Soderberg}, {Mattila}, {Kankare}, {Burgett}, {Chambers}, {Dombeck}, {Grav},
  {Heasley}, {Hodapp}, {Jedicke}, {Kaiser}, {Kudritzki}, {Luppino}, {Lupton},
  {Magnier}, {Monet}, {Morgan}, {Onaka}, {Price}, {Rhoads}, {Siegmund},
  {Stubbs}, {Tonry}, {Wainscoat}, {Waterson}, \& {Wynn-Williams}}]{Gezari10}
{Gezari} S. {et~al.}, 2010, ApJL, 720, L77

\bibitem[{{Hinton}, {Osindero} \& {Teh}(2006){Hinton}, {Osindero}, \&
  {Teh}}]{Hinton06}
{Hinton} G., {Osindero} S., {Teh} Y.~W., 2006, Neural computation, 18, 1527

\bibitem[{{Hodapp} {et~al}\mbox{.}(2004){Hodapp}, {Siegmund}, {Kaiser},
  {Chambers}, {Laux}, {Morgan}, \& {Mannery}}]{Hodapp04}
{Hodapp} K.~W., {Siegmund} W.~A., {Kaiser} N., {Chambers} K.~C., {Laux} U.,
  {Morgan} J., {Mannery} E., 2004, in Society of Photo-Optical Instrumentation
  Engineers (SPIE) Conference Series, Vol. 5489, Ground-based Telescopes,
  {Oschmann} Jr. J.~M., ed., pp. 667--678

\bibitem[{{Kaiser} {et~al}\mbox{.}(2010){Kaiser}, {Burgett}, {Chambers},
  {Denneau}, {Heasley}, {Jedicke}, {Magnier}, {Morgan}, {Onaka}, \&
  {Tonry}}]{Kaiser10}
{Kaiser} N. {et~al.}, 2010, in Society of Photo-Optical Instrumentation
  Engineers (SPIE) Conference Series, Vol. 7733, Society of Photo-Optical
  Instrumentation Engineers (SPIE) Conference Series

\bibitem[{{Keller} {et~al}\mbox{.}(2007){Keller}, {Schmidt}, {Bessell},
  {Conroy}, {Francis}, {Granlund}, {Kowald}, {Oates}, {Martin-Jones},
  {Preston}, {Tisserand}, {Vaccarella}, \& {Waterson}}]{Keller07}
{Keller} S.~C. {et~al.}, 2007, PASA, 24, 1

\bibitem[{{LeCun} {et~al}\mbox{.}(1998){LeCun}, {Bottou}, {Bengio}, \&
  {Haffner}}]{LeCun98}
{LeCun} Y., {Bottou} L., {Bengio} Y., {Haffner} P., 1998, Proceedings of the
  IEEE, 86, 2278

\bibitem[{{Lunnan} {et~al}\mbox{.}(2013){Lunnan}, {Chornock}, {Berger},
  {Milisavljevic}, {Drout}, {Sanders}, {Challis}, {Czekala}, {Foley}, {Fong},
  {Huber}, {Kirshner}, {Leibler}, {Marion}, {McCrum}, {Narayan}, {Rest},
  {Roth}, {Scolnic}, {Smartt}, {Smith}, {Soderberg}, {Stubbs}, {Tonry},
  {Burgett}, {Chambers}, {Kudritzki}, {Magnier}, \& {Price}}]{Lunnan13}
{Lunnan} R. {et~al.}, 2013, ApJ, 771, 97

\bibitem[{{Magnier}(2006)}]{Magnier06}
{Magnier} E., 2006, in The Advanced Maui Optical and Space Surveillance
  Technologies Conference

\bibitem[{{Magnier} {et~al}\mbox{.}(2013){Magnier}, {Schlafly}, {Finkbeiner},
  {Juric}, {Tonry}, {Burgett}, {Chambers}, {Flewelling}, {Kaiser}, {Kudritzki},
  {Morgan}, {Price}, {Sweeney}, \& {Stubbs}}]{Magnier13}
{Magnier} E.~A. {et~al.}, 2013, ApJS, 205, 20

\bibitem[{{McCrum} {et~al}\mbox{.}(2014){McCrum}, {Smartt}, {Kotak}, {Rest},
  {Jerkstrand}, {Inserra}, {Rodney}, {Chen}, {Howell}, {Huber}, {Pastorello},
  {Tonry}, {Bresolin}, {Kudritzki}, {Chornock}, {Berger}, {Smith},
  {Botticella}, {Foley}, {Fraser}, {Milisavljevic}, {Nicholl}, {Riess},
  {Stubbs}, {Valenti}, {Wood-Vasey}, {Wright}, {Young}, {Drout}, {Czekala},
  {Burgett}, {Chambers}, {Draper}, {Flewelling}, {Hodapp}, {Kaiser}, {Magnier},
  {Metcalfe}, {Price}, {Sweeney}, \& {Wainscoat}}]{McCrum14}
{McCrum} M. {et~al.}, 2014, MNRAS, 437, 656

\bibitem[{{Murtagh}(1991)}]{Murtagh91}
{Murtagh} F., 1991, Neurocomputing, 2, 183

\bibitem[{{Raina} {et~al}\mbox{.}(2007){Raina}, {Battle}, {Lee}, {Packer}, \&
  {Ng}}]{Raina07}
{Raina} R., {Battle} A., {Lee} H., {Packer} B., {Ng} A.~Y., 2007, in
  Proceedings of the 24th international conference on Machine learning, ACM,
  pp. 759--766

\bibitem[{{Rau} {et~al}\mbox{.}(2009){Rau}, {Kulkarni}, {Law}, {Bloom},
  {Ciardi}, {Djorgovski}, {Fox}, {Gal-Yam}, {Grillmair}, {Kasliwal}, {Nugent},
  {Ofek}, {Quimby}, {Reach}, {Shara}, {Bildsten}, {Cenko}, {Drake},
  {Filippenko}, {Helfand}, {Helou}, {Howell}, {Poznanski}, \&
  {Sullivan}}]{Rau09}
{Rau} A. {et~al.}, 2009, PASP, 121, 1334

\bibitem[{{Rest} {et~al}\mbox{.}(2014){Rest}, {Scolnic}, {Foley}, {Huber},
  {Chornock}, {Narayan}, {Tonry}, {Berger}, {Soderberg}, {Stubbs}, {Riess},
  {Kirshner}, J., {Schlafly}, {Rodney}, {Botticella}, {Brout}, {Challis},
  {Czekala}, {Drout}, {Hudson}, {Kotak}, {Leibler}, {Lunnan}, {Marion},
  {McCrum}, {Milisavljevic}, {Pastorello}, {Sanders}, {Smith}, {Stafford},
  {Thilker}, {Valenti}, {Wood-Vasey}, {Zheng}, {Burgett}, {Chambers},
  {Denneau}, {Draper}, {Flewelling}, {Hodapp}, {Kaiser}, {Kudritzki},
  {Magnier}, {Metcalfe}, {Price}, {Sweeney}, {Wainscoat}, \& {Waters}}]{Rest14}
{Rest} A. {et~al.}, 2014, ApJ, 795, 44

\bibitem[{{Rest} {et~al}\mbox{.}(2005){Rest}, {Stubbs}, {Becker}, {Miknaitis},
  {Miceli}, {Covarrubias}, {Hawley}, {Smith}, {Suntzeff}, {Olsen}, {Prieto},
  {Hiriart}, {Welch}, {Cook}, {Nikolaev}, {Huber}, {Prochtor}, {Clocchiatti},
  {Minniti}, {Garg}, {Challis}, {Keller}, \& {Schmidt}}]{Rest05}
{Rest} A. {et~al.}, 2005, ApJ, 634, 1103

\bibitem[{{Romano}, {Aragon} \& {Ding}(2006){Romano}, {Aragon}, \&
  {Ding}}]{Romano06}
{Romano} R.~A., {Aragon} C.~R., {Ding} C., 2006, in Machine Learning and
  Applications, 2006. ICMLA'06. 5th International Conference on, IEEE, pp.
  77--82

\bibitem[{{Saffari} {et~al}\mbox{.}(2009){Saffari}, {Leistner}, {Santner},
  {Godec}, \& {Bischof}}]{Saffari09}
{Saffari} A., {Leistner} C., {Santner} J., {Godec} M., {Bischof} H., 2009, in
  {Computer Vision Workshops (ICCV Workshops), 2009 IEEE 12th International
  Conference on}, pp. 1393--1400

\bibitem[{{Schlafly} {et~al}\mbox{.}(2012){Schlafly}, {Finkbeiner},
  {Juri{\'c}}, {Magnier}, {Burgett}, {Chambers}, {Grav}, {Hodapp}, {Kaiser},
  {Kudritzki}, {Martin}, {Morgan}, {Price}, {Rix}, {Stubbs}, {Tonry}, \&
  {Wainscoat}}]{Schlafly12}
{Schlafly} E.~F. {et~al.}, 2012, ApJ, 756, 158

\bibitem[{{Shalev-Shwartz}(2011)}]{Shalev-Shwartz11}
{Shalev-Shwartz} S., 2011, Foundations and Trends in Machine Learning, 4, 107

\bibitem[{{Smartt} {et~al}\mbox{.}(2014){Smartt}, {Smith}, {Wright}, {Young},
  {Kotak}, {Nicholl}, {Polshaw}, {Inserra}, {Chen}, {Terreran}, {Gall},
  {Fraser}, {McCrum}, {Valenti}, {Foley}, {Lawrence}, {Gezari}, {Burgett},
  {Chambers}, {Huber}, {Kudritzki}, {Magnier}, {Morgan}, {Tonry}, {Sweeney},
  {Stubbs}, {Kirshner}, {Metcalfe}, \& {Rest}}]{SmarttAtel14}
{Smartt} S.~J. {et~al.}, 2014, The Astronomer's Telegram, 6156, 1

\bibitem[{{Smartt} {et~al}\mbox{.}(2013){Smartt}, {Valenti}, {Fraser},
  {Inserra}, {Young}, {Sullivan}, {Benetti}, {Gal-Yam}, {Knapic}, {Molinaro},
  {Pastorello}, {Smareglia}, {Smith}, {Taubenberger}, \& {Yaron}}]{Smartt13}
{Smartt} S.~J. {et~al.}, 2013, The Messenger, 154, 50

\bibitem[{{Tonry} {et~al}\mbox{.}(2012{\natexlab{a}}){Tonry}, {Stubbs},
  {Kilic}, {Flewelling}, {Deacon}, {Chornock}, {Berger}, {Burgett}, {Chambers},
  {Kaiser}, {Kudritzki}, {Hodapp}, {Magnier}, {Morgan}, {Price}, \&
  {Wainscoat}}]{Tonry12a}
{Tonry} J.~L. {et~al.}, 2012{\natexlab{a}}, ApJ, 745, 42

\bibitem[{{Tonry} {et~al}\mbox{.}(2012{\natexlab{b}}){Tonry}, {Stubbs},
  {Lykke}, {Doherty}, {Shivvers}, {Burgett}, {Chambers}, {Hodapp}, {Kaiser},
  {Kudritzki}, {Magnier}, {Morgan}, {Price}, \& {Wainscoat}}]{Tonry12b}
{Tonry} J.~L. {et~al.}, 2012{\natexlab{b}}, ApJ, 750, 99

\bibitem[{{York} {et~al}\mbox{.}(2000){York}, {Adelman}, {Anderson},
  {Anderson}, {Annis}, {Bahcall}, {Bakken}, {Barkhouser}, {Bastian}, {Berman},
  {Boroski}, {Bracker}, {Briegel}, {Briggs}, {Brinkmann}, {Brunner}, {Burles},
  {Carey}, {Carr}, {Castander}, {Chen}, {Colestock}, {Connolly}, {Crocker},
  {Csabai}, {Czarapata}, {Davis}, {Doi}, {Dombeck}, {Eisenstein}, {Ellman},
  {Elms}, {Evans}, {Fan}, {Federwitz}, {Fiscelli}, {Friedman}, {Frieman},
  {Fukugita}, {Gillespie}, {Gunn}, {Gurbani}, {de Haas}, {Haldeman}, {Harris},
  {Hayes}, {Heckman}, {Hennessy}, {Hindsley}, {Holm}, {Holmgren}, {Huang},
  {Hull}, {Husby}, {Ichikawa}, {Ichikawa}, {Ivezi{\'c}}, {Kent}, {Kim},
  {Kinney}, {Klaene}, {Kleinman}, {Kleinman}, {Knapp}, {Korienek}, {Kron},
  {Kunszt}, {Lamb}, {Lee}, {Leger}, {Limmongkol}, {Lindenmeyer}, {Long},
  {Loomis}, {Loveday}, {Lucinio}, {Lupton}, {MacKinnon}, {Mannery}, {Mantsch},
  {Margon}, {McGehee}, {McKay}, {Meiksin}, {Merelli}, {Monet}, {Munn},
  {Narayanan}, {Nash}, {Neilsen}, {Neswold}, {Newberg}, {Nichol}, {Nicinski},
  {Nonino}, {Okada}, {Okamura}, {Ostriker}, {Owen}, {Pauls}, {Peoples},
  {Peterson}, {Petravick}, {Pier}, {Pope}, {Pordes}, {Prosapio},
  {Rechenmacher}, {Quinn}, {Richards}, {Richmond}, {Rivetta}, {Rockosi},
  {Ruthmansdorfer}, {Sandford}, {Schlegel}, {Schneider}, {Sekiguchi}, {Sergey},
  {Shimasaku}, {Siegmund}, {Smee}, {Smith}, {Snedden}, {Stone}, {Stoughton},
  {Strauss}, {Stubbs}, {SubbaRao}, {Szalay}, {Szapudi}, {Szokoly}, {Thakar},
  {Tremonti}, {Tucker}, {Uomoto}, {Vanden Berk}, {Vogeley}, {Waddell}, {Wang},
  {Watanabe}, {Weinberg}, {Yanny}, {Yasuda}, \& {SDSS Collaboration}}]{York00}
{York} D.~G. {et~al.}, 2000, AJ, 120, 1579

\end{thebibliography}

\begin{table*}
    \begin{minipage}{140mm} 
    \caption{First detections of the 53
    spectroscopically confirmed PS1 supernovae ordered by hypothesis.
    (Due to sky cells some
    SNe appear twice.)} 
    \begin{tabular}{lccccc}
    \hline Name   & Classification & First Detection (MJD) &  Magnitude & Filter & Hypothesis\\ 
    \hline \hline 
    PS1-13duq & Ia & 56588.381 & 21.64 & i$_{P1}$ &0.989 \\
    PS1-13bzp & Ia & 56478.276 & 21.33 & g$_{P1}$ & 0.98 \\ 
    PS1-13bzp & Ia & 56478.276 & 21.30 & g$_{P1}$ & 0.975 \\
    PS1-13abg & II-P & 56383.336 & 21.47 & z$_{P1}$ & 0.971 \\
    PS1-13bqg & Ia & 56443.287 & 20.90 & g$_{P1}$ & 0.968 \\ 
    PS1-13abg & II-P & 56383.336 & 21.55 & z$_{P1}$ & 0.963 \\ 
    PS1-14il & IIn & 56676.554 & 21.36 & z$_{P1}$ & 0.959 \\ 
    PS1-13vc & Ia & 56351.476 & 20.67 & z$_{P1}$ & 0.958 \\ 
    PS1-13abw & Ic & 56383.438 & 21.23 & z$_{P1}$ & 0.958 \\
    PS1-14ky & II & 56681.499 & 21.60 & z$_{P1}$ & 0.957 \\ 
    PS1-13ur & Ia & 56351.525 & 20.32 & z$_{P1}$ & 0.955 \\ 
    PS1-13eae & II & 56604.598 & 19.82 & y$_{P1}$ & 0.954 \\
    PS1-13alz & II-P & 56399.282 & 20.43 & i$_{P1}$ & 0.952 \\
    PS1-12cnr & Ia & 56283.340 & 20.33 & z$_{P1}$ & 0.948 \\
    PS1-13can & Ia & 56477.567 & 22.04 & z$_{P1}$ & 0.946 \\
    PS1-13cws & Ia & 56549.460 & 21.48 & z$_{P1}$ & 0.943 \\
    PS1-13ge & Ia & 56328.612 & 21.49 & g$_{P1}$ & 0.94 \\ 
    PS1-12cho & Ia & 56262.469 & 21.07 & z$_{P1}$ & 0.933 \\
    PS1-12cey & II & 56268.294 & 22.06 & g$_{P1}$ & 0.931 \\
    PS1-13bok & I & 56424.560 & 22.35 & r$_{P1}$ & 0.926 \\ 
    PS1-13djz & Ic & 56554.585 & 20.74 & z$_{P1}$ & 0.923 \\
    PS1-13a & Ia & 56289.280 & 21.20 & z$_{P1}$ & 0.919 \\ 
    PS1-13bit & Ia & 56420.548 & 22.69 & i$_{P1}$ & 0.918 \\ 
    PS1-13bqb & Ia & 56443.287 & 22.32 & g$_{P1}$ & 0.918 \\
    PS1-13djj & Ia & 56563.576 & 20.72 & g$_{P1}$ & 0.916 \\
    PS1-12bza & II-P & 56262.469 & 21.12 & z$_{P1}$ & 0.914 \\
    PS1-13brf & Ia & 56443.324 & 22.68 & r$_{P1}$ & 0.91 \\ 
    PS1-13hp & II-P & 56325.545 & 21.40 & g$_{P1}$ & 0.907 \\
    PS1-13adg & Ia &   56384.515 & 21.71 & r$_{P1}$ & 0.902 \\
    PS1-13awf & I & 56417.315 & 22.47 & i$_{P1}$ & 0.899 \\ 
    PS1-13atm & II-P & 56410.298 & 22.25 & z$_{P1}$ & 0.898 \\
    PS1-13cjb & II & 56501.436 & 22.63 &g$_{P1}$ & 0.896 \\ 
    PS1-12cho & Ia & 56262.469 & 21.08 & z$_{P1}$ & 0.884 \\
    PS1-13cai & Ia & 56477.567 & 21.99 & z$_{P1}$ & 0.882 \\
    PS1-13bni &II-P & 56420.548 & 23.07 & i$_{P1}$ & 0.873 \\
    PS1-13bog & Ia & 56417.341 & 22.57 & i$_{P1}$ & 0.867 \\  
    PS1-12chw & Ia & 56262.313 & 21.21 & y$_{P1}$ & 0.857 \\
    PS1-13bqv & Ia & 56442.486 & 21.64 & z$_{P1}$ & 0.849 \\
    PS1-13djs & Ia & 56562.587 & 21.44 & z$_{P1}$ & 0.84 \\ 
    PS1-13aai/SN 2013au & Ia & 56370.421 & 19.29 & z$_{P1}$ & 0.833 \\ 
    PS1-13cuc/SN 2013go & Ia & 56536.587 & 19.03 & z$_{P1}$ & 0.814 \\
    PS1-13hs & I & 56328.515 & 21.97 & g$_{P1}$ & 0.801 \\
    PS1-13baf & II-P &56414.521 & 22.54 & i$_{P1}$ & 0.801 \\
    PS1-13avb & Ib & 56414.521 & 21.75 & i$_{P1}$ & 0.796 \\ 
    PS1-13ayn & Ia & 56416.449 & 22.01 & r$_{P1}$ & 0.77 \\ 
    PS1-13aai/SN 2013au & Ia & 56370.421 & 19.36 & z$_{P1}$ & 0.735 \\ 
    PS1-13fo/SN 2013X & Ia & 56314.625 & 18.04 & y$_{P1}$ & 0.713 \\ 
    PS1-13bus & Ia & 56462.440 & 22.91 & i$_{P1}$ & 0.708 \\
    PS1-13brw & II-P & 56436.325 & 20.71 & y$_{P1}$ & 0.707 \\ 
    PS1-13hi & IIn & 56324.604 & 18.50 & z$_{P1}$ & 0.704 \\
    PS1-13bvc & Ia & 56469.349 & 21.90 & z$_{P1}$ & 0.698 \\
    PS1-13bzk & Ia & 56468.571 & 21.38 & y$_{P1}$ & 0.662 \\ 
    PS1-13abf & Ia & 56380.368 & 20.21 & y$_{P1}$ & 0.642 \\ 
    PS1-13arv & Ia & 56409.239 & 20.98 & y$_{P1}$ & 0.638 \\
    PS1-13wr & II-P & 56349.602 & 20.39 & y$_{P1}$ & 0.637 \\  
    PS1-14xz/SN 2014bc & II-P & 56399.380 & 18.25 & i$_{P1}$ & 0.623\\ 
    PS1-13wr & II-P & 56349.602 & 20.40 & y$_{P1}$ & 0.61 \\
    PS1-13buf & Ia & 56461.299 & 22.60 & z$_{P1}$ & 0.577 \\
    \hline 
    \end{tabular}
    \label{tab:firstDetections} 
    \end{minipage}
     \end{table*}

\end{document}